\documentclass[aps,prd,10pt,nofootinbib,twocolumn,superscriptaddress,floatfix,notitlepage]{revtex4-1}
\pdfoutput=1

\usepackage{graphicx}
\usepackage{amsmath,amsfonts,amssymb}
\usepackage[dvipsnames]{xcolor}
\usepackage[breaklinks,colorlinks,urlcolor=MidnightBlue,citecolor=WildStrawberry,linkcolor=Fuchsia]{hyperref}
\usepackage{verbatim}
\usepackage{enumitem}
\usepackage{aas_macros}

\newcommand{\Msun}{M_\odot}
\newcommand{\td}{{\rm d}}

\newcommand{\be}{\begin{equation}}
\newcommand{\ee}{\end{equation}}
\newcommand{\bea}{\begin{equation} \begin{aligned}}
\newcommand{\eea}{\end{aligned} \end{equation}}

\def\lsim{\mathrel{\raise.3ex\hbox{$<$\kern-.75em\lower1ex\hbox{$\sim$}}}}
\def\gsim{\mathrel{\raise.3ex\hbox{$>$\kern-.75em\lower1ex\hbox{$\sim$}}}}

\begin{document}

\title{Statistics of the supermassive black hole gravitational wave background anisotropy}

\author{Juhan Raidal}
\email{juhan.raidal@kbfi.ee}
\affiliation{Laboratory of High Energy and Computational Physics, NICPB, R{\"a}vala 10, Tallinn, 10143, Estonia}
\affiliation{Departament of Cybernetics, Tallinn University of Technology, Akadeemia tee 21, 12618 Tallinn, Estonia}

\author{Juan Urrutia}
\email{juan.urrutia@kbfi.ee}
\affiliation{Laboratory of High Energy and Computational Physics, NICPB, R{\"a}vala 10, Tallinn, 10143, Estonia}
\affiliation{Departament of Cybernetics, Tallinn University of Technology, Akadeemia tee 21, 12618 Tallinn, Estonia}

\author{Ville Vaskonen}
\email{ville.vaskonen@pd.infn.it}
\affiliation{Laboratory of High Energy and Computational Physics, NICPB, R{\"a}vala 10, Tallinn, 10143, Estonia}
\affiliation{Dipartimento di Fisica e Astronomia, Universit\`a degli Studi di Padova, Via Marzolo 8, 35131 Padova, Italy}
\affiliation{Istituto Nazionale di Fisica Nucleare, Sezione di Padova, Via Marzolo 8, 35131 Padova, Italy}

\author{Hardi Veerm\"ae}
\email{hardi.veermae@cern.ch}
\affiliation{Laboratory of High Energy and Computational Physics, NICPB, R{\"a}vala 10, Tallinn, 10143, Estonia}

\begin{abstract}
We study the statistical properties of the anisotropy in the gravitational wave (GW) background originating from supermassive black hole (SMBH) binaries. We derive the distribution of the GW anisotropy power spectrum coefficients, $C_{l\geq1}/C_0$, in scenarios including environmental effects and eccentricities of the SMBH binaries. Although the mean of $C_{l\geq1}/C_0$ is the same for all multipoles, we show that their distributions vary, with the low $l$ distributions being the widest. We also find a strong correlation between spectral fluctuations and the anisotropy in the GW signal. We show that the GW anisotropy can break the degeneracy between the scenarios including environmental effects or eccentricity. In particular, we find that existing NANOGrav constraints on GW anisotropy begin to constrain SMBH scenarios with strong environmental effects.
\end{abstract}

\maketitle


\section{Introduction}

Pulsar timing array (PTA) collaborations have provided compelling evidence for the existence of a stochastic gravitational wave (GW) background by measuring both a common-spectrum stochastic process~\cite{NANOGrav:2023gor, EPTA:2023fyk, EPTA:2023sfo, Reardon:2023gzh, Zic:2023gta, Reardon:2023zen, Xu:2023wog} and finding strong evidence for Hellings-Downs angular correlations~\cite{Hellings:1983fr}. The most conservative explanation for the origin of the background is supermassive black hole (SMBH) binaries~\cite{EPTA:2023xxk, NANOGrav:2023hde, Ellis:2023dgf, Bi:2023tib, Ellis:2024wdh, Raidal:2024odr}, although cosmological sources, such as cosmic strings, phase transitions and domain walls, can explain the data equally well~\cite{EPTA:2023xxk, NANOGrav:2023hvm, Figueroa:2023zhu, Ellis:2023oxs}. Unlike a GW background originating from SMBHs, cosmological sources generate an isotropic background, making the detection of anisotropy within the GW background measurements a powerful indicator for SMBH binaries being the source of the background. 

Methods for measuring GW background anisotropies using PTAs have been well-explored~\cite{Mingarelli:2013dsa, Cornish:2014rva, Allen:2024mtn, Hotinli:2019tpc} and investigations into simulated circular SMBH binary populations have explored both the magnitude and frequency scaling of anisotropies in these models~\cite{Taylor:2013esa, Taylor:2020zpk, Gardiner:2023zzr, Sato-Polito:2023spo, Sah:2024oyg}. With the most recent constraint on measured anisotropy placed at $C_{l>0}/C_{l=0} < 0.2$ for the Bayesian 95\% upper limit by NANOGrav~\cite{NANOGrav:2023tcn}, we can already derive information about SMBH binary populations based on the amount of anisotropy they produce in the GW signal. Current and future experiments will, within relatively short time frames as detailed by~\cite{Pol:2022sjn, Lemke:2024cdu, Depta:2024ykq}, significantly improve the sensitivity to anisotropies, enabling further model discrimination.

The distribution of fluctuations in the GW spectrum is dominated by a relatively small number of loud SMBH binaries~\cite{Ellis:2023owy, Ellis:2023dgf, Sato-Polito:2024lew, Xue:2024qtx}. Similarly, the loudest binaries generate a significant anisotropy in the background. In this paper, we aim to outline the statistical properties of GW background anisotropies by deriving the distributions of the angular power spectrum coefficients. Although the mean of the coefficients $C_{l>0}/C_{l=0}$ is independent of $l$, their distributions are not. As we will show, the distributions are highly non-Gaussian and can be quite wide. Thus, to make reliable inferences about the SMBH population based on GW anisotropy measurements one cannot rely purely on the mean of $C_{l>0}/C_{l=0}$. 

Processes that govern the orbital decay of the SMBH binaries and lead to their eventual merger are poorly understood. To reproduce the spectral shape of the GW background, either environmental effects or high eccentricities are needed~\cite{EPTA:2023xxk, NANOGrav:2023hde, Ellis:2023dgf, Bi:2023tib, Raidal:2024odr}. Although the timescale for the evolution of a single binary is similar in both scenarios, the physical process is very different. Binary evolution can be driven by its interactions with the surrounding matter before the GW emission takes over at very small separations. This reduces the amplitude of the GW background at lower frequencies which is emitted by wider binaries. On the other hand, highly eccentric binaries evolve faster at low orbital frequencies by emitting GWs at multiple harmonics, consequently contributing to the GW background also at higher frequencies~\cite{Enoki:2006kj,Sesana:2013wja}. This results in a spectrum that is similar to the one obtained with environmental energy loss. However, as we will show, the GW background from eccentric binaries is significantly more isotropic due to a larger number of binaries contributing to it in a given frequency range. Therefore, the GW background anisotropy is an observable that can discriminate between not only cosmological and astrophysical sources but also between the models SMBH binary evolution.

\begin{figure*}[t]
    \centering  
    \includegraphics[width=\textwidth]{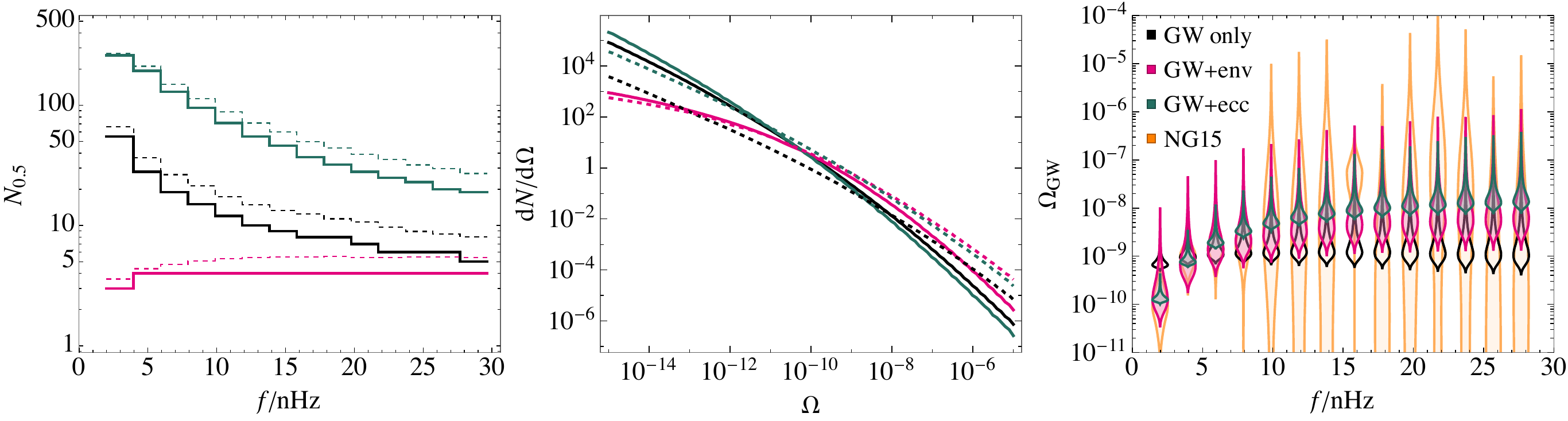} 
    \caption{GW characteristics of the SMBH binary population for three different SMBH models: purely GW driven SMBH inspirals (GW only), SMBH inspirals with environmental effects (GW+env), and inspirals of eccentric SMBH binaries (GW+ecc). \emph{Left panel:} The median (solid line) and the mean (dashed line) number of the strongest sources that make up half the signal at each PTA frequency band. \emph{Middle panel:} The distributions of the contributions from individual binaries at $1.98$\,nHz (solid lines) and $27.68$\,nHz (dashed lines). \emph{Right panel:} The distributions of the GW spectra in the $n\times1.98$\,nHz frequency bins compared to the NG15 year data (yellow violins).} 
    \label{fig:bestfit}
\end{figure*}

This paper is organized as follows: In section~\ref{sec:models}, we will review the SMBH models considered here and the computation of the GW background from a given SMBH scenario. The GW background anisotropy and its statistical properties are derived in section~\ref{sec:anisotropy}. Section~\ref{sec:results} summarizes the main results and their implications are discussed in section~\ref{sec:discussion}. We conclude in section~\ref{sec:concl}. Some technical results are gathered in the appendices.

\section{Modelling the SMBH population}
\label{sec:models}

Although the SMBHs have been found at the centres of many galaxies, their formation mechanisms and evolution are presently not well understood. If the PTA signal is to be interpreted as coming from SMBH binaries, different merger rates and evolutions of binaries can be contrasted to the observation. In this paper, we study three models that describe the GW emission of SMBH binaries. The first two models consider circular binaries, evolving purely through GW emission in the first model and via GW emission and environmental interactions in the second~\cite{Ellis:2023dgf}. The third model considers eccentric binaries evolving purely through GW emission~\cite{Raidal:2024odr}. Their GW-related characteristics are summarized in Fig.~\ref{fig:bestfit}.

The GW background of SMBHs is determined by their merger rate. The differential merger rate of BHs in the observer reference frame is\footnote{Geometric units $c = G = 1$ are used throughout the paper.}
\be \label{eq:diffmergerrate}
    \td \lambda = \td \mathcal{M} \td \eta \td z \frac{1}{1+z} \frac{\td V_c}{\td z} \frac{\td R_{\rm BH}}{\td \mathcal{M} \td \eta}\,,
\ee
where $\mathcal{M}$, $\eta$ and $z$ are chirp mass, the symmetric mass ratio and the redshift of the binary, $V_c$ is the comoving volume, and $R_{\rm BH}$ is the comoving BH merger rate density. We use the estimate of the BH merger rate coming from the halo merger rate $R_h$ given by the extended Press-Schechter formalism~\cite{Press:1973iz,Bond:1990iw,1993MNRAS.262..627L}:
\bea \label{eq:mergerrate}
    \frac{\td R_{\rm BH}}{\td m_1 \td m_2} = \,p_{\rm BH} \int & \td M_{{\rm v},1} \td M_{{\rm v},2} \, \frac{\td R_h}{\td M_{{\rm v},1} \td M_{{\rm v},2}}\\
    & \times \prod_{j=1,2} \frac{\td P(m_j|M_*(M_{{\rm v},j},z))}{\td m_j}  \,, 
\eea
where $m_{j}$ are the masses of the merging BHs, $M_{{\rm v},j}$ are the virial masses of their host halos and $p_{\rm BH} \le 1$, treated as a free parameter, combines the SMBH occupation fraction in galaxies with the efficiency for the BHs to merge following the merger of their host halos. The BH masses are obtained using the observed halo mass-stellar mass relation $M_*(M_{{\rm v},j},z)$ taken from~\cite{Girelli:2020goz} and the global fit stellar mass-BH mass relation~\cite{Ellis:2024wdh}
\be\label{eq:MH_relation}
    \frac{\td P(m|M_*)}{\td \log_{10} \!m} = \mathcal{N}\bigg(\!\log_{10} \!\frac{m}{M_\odot} \bigg| a + b \log_{10} \!\frac{M_*}{10^{11}M_\odot},\sigma\bigg) \, ,
\ee
with $a = 8.6$, $b = 0.8$ and $\sigma = 0.8$, where $\mathcal{N}(x|\bar x,\sigma)$ denotes the probability density of a Gaussian distribution with mean $\bar x$ and variance $\sigma^2$.

The mean of the GW background spectrum at a given frequency from a population of potentially eccentric SMBH binaries, in its most general form, is given by~\cite{Raidal:2024odr,Ellis:2023dgf}
\bea \label{eq:mean}
    \langle\Omega(f)\rangle = \frac{1}{\rho_c} \int \td \lambda\,& \td e\, \td f_{\rm b}\, \sum_{n=1}^\infty \delta\!\left(f_{\rm b} - \frac{f_{\rm r}}{n}\right) \\ 
    &\times  P(e|f_{\rm b})  \frac{1+z}{4\pi D_L^2} \frac{\td E_n}{\td \ln{f_{\rm b}}} \,,
\eea
where $\rho_c$ is the critical energy density, $D_L$ is the source luminosity distance, $P(e|f_{\rm b})$ is the binary eccentricity distribution at a given orbital frequency $f_{\rm b}$ and the delta function picks binaries whose $n$-th harmonic contributes to the observed frequency $f = f_{\rm r}/(1+z)$. We model $P(e|f_{\rm b})$ by a power law (see \cite{Raidal:2024odr}) for the eccentric model and $P(e|f_{\rm b}) = \delta(e)$ for the models with circular binaries. Expressing Eq.~\eqref{eq:mean} in terms of binary parameters gives
\bea \label{eq:mean2}
    \langle\Omega(f)\rangle 
    &\!= \!\!\int \!\! \td \lambda \td e \sum_{n=1}^\infty P(e|f_{\rm b}) g_n(e) \Omega_c^{(1)}\! \frac{\td t}{\td \ln{f_{\rm b}}} \bigg|_{f_{\rm b}\!=\!\frac{f_r}{n}}\!\!,
\eea
where $g_n(e)$ is the GW power emitted into the $n$-th harmonic \cite{Peters:1963ux}, $\td t/\td \ln{f_{\rm b}}$ is the residence time that is affected by environment and eccentricity of the binary (see Appendix~\ref{app:ecc} for details) and
\be\label{eq:Omega1}
    \Omega_c^{(1)} \equiv \frac{1}{4\pi D_L^2\rho_c}\frac{\td E_c}{\td t_r}
\ee
gives the GW luminosity of a single circular binary. 

It is well known that the mean~\eqref{eq:mean2} does not capture the likely realizations of the GW spectrum well, because the distributions of the spectral fluctuations are not Gaussian and possess a heavy tail favouring large upward fluctuations~\cite{Ellis:2023owy, Ellis:2023dgf, Sato-Polito:2024lew, Xue:2024qtx}. Therefore, it is important to consider the statistical fluctuations in the signal at a given measured frequency. These statistical properties can be inferred from the distribution of contributions $\Omega$ from individual binaries at a given frequency $f$:
\bea \label{eq:P1semi-analytic}
    P^{(1)}(\Omega|f) 
    =& \left( \frac{\td N}{\td \ln f} \right)^{-1} \int \td e \,\td\lambda \sum_{n=1}^\infty P(e|f_{\rm b}) \\
    &\times \frac{\td t}{\td\ln{f_{\rm b}}} \delta(\Omega - \Omega_c^{(1)} g_n(e)) \bigg|_{f_{\rm b} = \frac{f_r}{n}} \,,
\eea
where 
\be
    \frac{\td N}{\td \ln f} = \int \td \lambda \frac{\td t}{\td \ln f_{\rm b}}\bigg|_{f_{\rm b} = \frac{f_{r}}{2}} \,
\ee
is the number of binaries per logarithmic frequency interval.

\begin{figure*}
        \centering
        \includegraphics[width= 0.85\paperwidth]{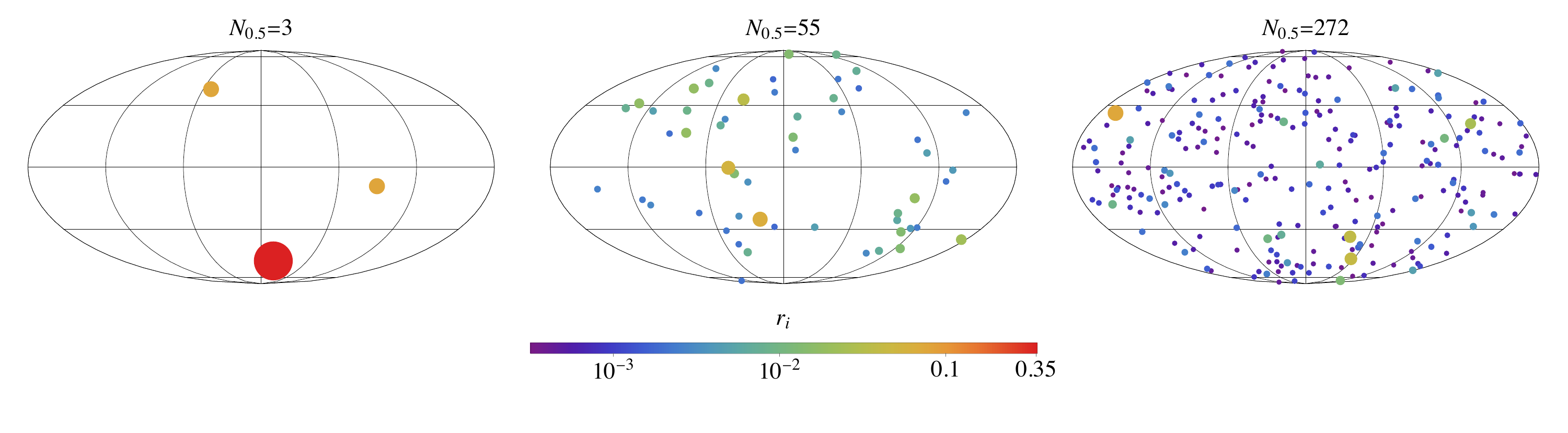}
        \vspace{-8mm}
        \caption{Typical realization of the strongest binaries contributing 50\% of the GW signal in the $1.98$\,nHz bin for the GW+env (left), GW-only (middle) and GW+ecc (right) models. The colour and size of the binaries indicate the relative magnitude of GW energy emitted from the binary.}
        \label{fig:skymap}
\end{figure*}

Given a model for the SMBH binary population, the spectral distributions of the total GW abundance in each frequency bin, $P(\Omega|f_j)$, can be obtained by combining Monte Carlo (MC) sampling for louder binaries (i.e. exceeding a threshold $\Omega > \Omega_{\rm thr}$) and analytic estimates for the remaining quieter binaries (for details, see~\cite{Ellis:2023dgf, Raidal:2024odr}). We will choose the threshold in a way that the number of strong binaries in each frequency bin is $N_S=50$. This number of strong sources was found in~\cite{Ellis:2023dgf} to be sufficient to accurately model the distribution of the high-amplitude fluctuations in the GW spectrum whilst balancing the computational cost.

In our numerical analysis, we will consider three models with qualitatively different SMBH binary dynamics:
\begin{enumerate}[leftmargin=*]
    \item {\bf GW only}: circular binaries evolving by GW emission only ($p_{\rm BH} = 0.06$),
    \item {\bf GW+env}: circular binaries with energy loss by environmental interactions ($p_{\rm BH} = 0.37$, $f_{\rm ref} = 30$\,nHz),
    \item {\bf GW+ecc}: eccentric binaries evolving by GW emission only ($p_{\rm BH} = 0.60$, $\langle e\rangle_{2\,{\rm nHz}} = 0.83$).
\end{enumerate}
For each model, we fix the model parameters to the NANOGrav 15-year data (NG15) best-fit values found in Refs.~\cite{Ellis:2024wdh, Raidal:2024odr}. These values are shown in brackets in the above list (see Appendix~\ref{app:ecc} for the parametrization of the eccentricity distribution and the environmental effects).

The left panel of Fig.~\ref{fig:bestfit} shows the number of the strongest contributions making up half of the signal at each PTA frequency band. Note that eccentric binaries can contribute to multiple frequency bins, which is the main cause for the larger number of strong sources. The distribution of GW luminosities from individual binaries in a frequency range $[f_i, f_{i+1}]$ (the $i$-th frequency bin) is given by
\be\label{eq:dNdOmega}
    \frac{\td N_i}{\td \Omega} 
    = \int^{f_{i+1}}_{f_i} \td \ln f \, P^{(1)}(\Omega|f) \frac{\td N (f)}{\td \ln f}
\ee
The middle panel of Fig.~\ref{fig:bestfit} shows the GW luminosity functions in the $1.98$\,nHz and the $27.68$\,nHz bins. Compared to the GW only case, the inclusion of environmental effects suppresses the number of quiet sources, while the model containing eccentric binaries predicts relatively similar shapes, but a larger number of sources at higher frequencies. This occurs because the eccentricity decreases as the binaries shrink due to GW emission. We remark that, since the parameters are chosen via best fits, the merger efficiency $p_{\rm BH}$ in the merger rate~\eqref{eq:mergerrate} also varies between these models. Because of this, the overall normalization of the GW luminosity function does not arise purely due to differences in orbital dynamics.

The resulting GW spectra in the three models are shown in the right panel of Fig.~\ref{fig:bestfit}. Since the spectra depend on the random realization of the SMBH population in a given model, they are represented by violins characterising the distribution of spectral fluctuations $P(\Omega|f)$ at each frequency bin. Compared to the model of purely GW-driven circular binaries, environmental effects and binary eccentricities reduce the residence time at low frequencies and suppress the GW signal thus improving the fit to the PTA data.

\section{GW background anisotropy}
\label{sec:anisotropy}

The GW signal from a given direction $\hat{x}$ at a given frequency bin is given by\footnote{On the sphere, the Dirac delta function should be understood as $\delta^2(\hat{x} - \hat{x}_i) = \delta(\cos(\theta) - \cos(\theta_i))\delta(\varphi - \varphi_i)$ in terms of the inclination and the azimuth.}
\be
    \Omega(\hat{x}) = \sum_{i} \Omega_i \,\delta^2(\hat{x} - \hat{x}_i)\, ,
\ee
where $\Omega_i$ and $\hat{x}_i = (\sin\theta_i \cos\phi_i, \sin\theta_i \sin\phi_i,\cos\theta_i)$ denote the strong sources' signal strength and sky location. Expanding in spherical harmonics,
\be
    \Omega(\hat{x}) = \sum_{l=0}^\infty \sum_{m=-l}^{l}c_{lm}Y_{lm}(\theta,\phi) \,,
\ee
we have, for each realization, that
\bea\label{eq:clm}
    c_{lm}
    &= \int \td^2 \hat{x} \, \Omega(\hat{x}) Y^*_{lm}(\theta,\phi) 
    = \sum_{i} \Omega_i Y^{*}_{lm}(\theta,\phi) \,.
\eea
The anisotropy present in the spectrum is characterised by the power spectrum coefficients
\be\label{eq:Cl}
    C_l = \frac{1}{2l+1}\sum_{m=-l}^l |c_{lm}|^2,
\ee
which represent the measure of fluctuations in the power spectrum on the angular scale $\theta \approx 2\pi/l$. For a single source $C_l = \Omega_i^2/(4\pi)$ while for a large number of equally strong sources distributed uniformly on the sky $C_{l>0} \approx 0$.

The first coefficient $C_0 = \Omega_{\rm tot}^2/(4\pi)$, where $\Omega_{\rm tot} \equiv \sum_i \Omega_i$, represents the isotropic contribution and gives the normalization of the higher multipoles. To factor out its contribution, we will consider the multipole ratio
\be\label{clc0}
    \frac{C_{l}}{C_0} 
    = \sum_{i,j} r_i r_j P_{l}(\hat x_i\cdot \hat x_j) \,,
\ee
where 
$
    r_i \equiv \Omega_i/\Omega_{\rm tot}
$
is the fractional energy density of each binary, $P_l$ are Legendre polynomials and we used the addition theorem for spherical harmonics to simplify the angular dependence. Note that $0 \leq C_{l \geq 1}/C_0 \leq 1$ since $\sum_i r_i = 1$ and $|P_l| \leq 1$.

To model the GW signal, we will assume that binary sky locations are statistically isotropic and independent of the signal strength. Deviations from statistical isotropy due to large-scale inhomogeneities of matter can be relevant, but only for very high multipoles~\cite{Semenzato:2024mtn} which we will not consider here. Fig.~\ref{fig:skymap} shows example realizations of skymaps containing the strongest sources constituting half the signal in the $1.98$\,nHz band. Each realization corresponds to one of the three SMBH models described in Sec.~\ref{sec:models} with the $N_{0.5}$ close to its median value as given in the left panel of Fig.~\ref{fig:bestfit}.

Consider now the leading statistical momenta of the anisotropy coefficients. The angular average gives $\langle P_{l}(\hat x_i\cdot \hat x_j) \rangle = \delta_{ij}$, so the mean multipole ratios are independent of $l$,
\bea\label{eq:ClMean}
    \bigg\langle \frac{C_{l}}{C_0} \bigg\rangle 
    &= \sum_i \left\langle r_i^2\right\rangle\,,
    \qquad l \geq 1\,.
\eea
Such an $l$-independence, does, however, not hold for higher moments. An analogous computation gives the variance (for the derivation, see Appendix~\ref{app:stat})
\be\label{eq:sigma}
    \sigma_{C_l/C_0}^2
    \!=\! \sum_{i,j} \left(\langle r_i^2 r_j^2 \rangle \!-\! \langle r_i^2 \rangle \langle r_j^2 \rangle\right) 
    \!+\! \frac{2}{2l+1} \! \sum_{i\neq j}\langle r_i^2 r_j^2\rangle.
\ee
Although it shows $l$-dependence in the lower multipoles, the $l$-dependent terms are suppressed as $l$ grows. As we show below, this is a general property of all expectation values. Deviations from this $1/(2l+1)$ suppression would indicate the presence of statistical anisotropy.

\subsection{Large $l$ limit} 

\begin{figure}
    \centering    
    \includegraphics[width=0.95\columnwidth]{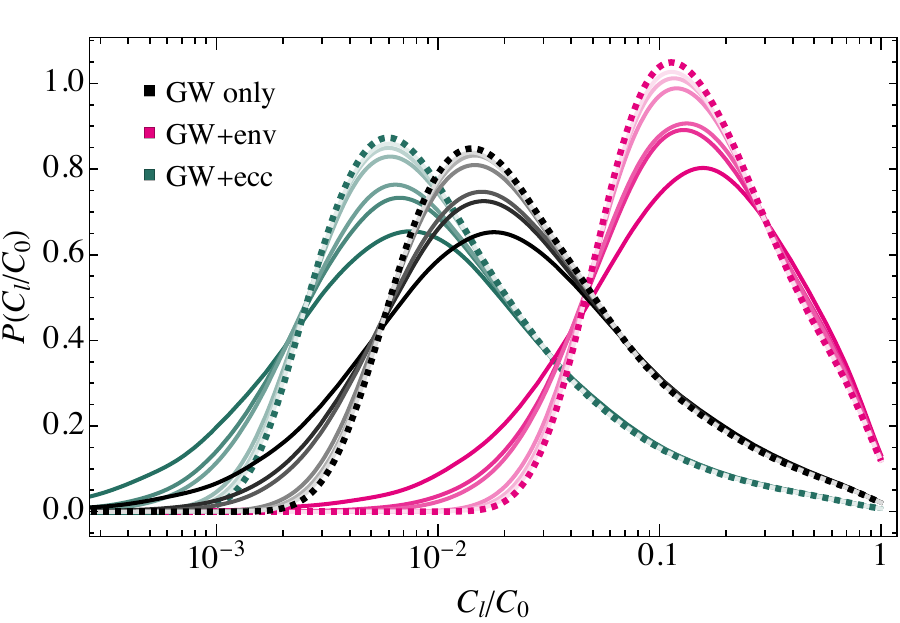}
    \vspace{-2mm}
    \caption{The solid curves show the distributions of the multipole ratios for $l \in\{ 1,2,3,10,20,40 \}$, with a lighter colour indicating a higher multipole, in the three models at the first PTA frequency band. The dashed curve shows the distribution of the sum of the fractional energy density squared of all the GW emitting sources, $C_{\infty}/C_0$.}
    \label{fig:PdfClC0}
\end{figure}

As demonstrated in Fig.~\ref{fig:PdfClC0}, we find that for $l\gg 1$, the fluctuations of the multipole ratios $C_{l}/C_0$ converge statistically to
\be\label{eq:C_infty}
    \lim_{l\to \infty} \frac{C_{l}}{C_0} = \sum_{i} r_i^2 \equiv \frac{C_{\infty}}{C_0} \,.
\ee
This can be shown by considering the fluctuations of the difference
\be
    \frac{C_{l}}{C_0} - \frac{C_{\infty}}{C_0} = 2\sum_{i>j} r_i r_j P_{l}(\hat x_i\cdot \hat x_j) \,.
\ee
The mean of this difference vanishes, $\left\langle C_{l}/C_{0} - C_{\infty}/C_{0} \right\rangle = 0$, and thus its variance is given by 
\be\label{eq:sigma_C_infty}
    \left\langle \left(\frac{C_{l}}{C_0} - \frac{C_{\infty}}{C_0}\right)^2 \right\rangle 
    = 
    \frac{2}{2l+1}\sum_{i\neq j} \left\langle r_i^2 r_j^2 \right\rangle\,. 
\ee
We see that the quantities $C_{l}/C_{0}$ and $C_{\infty}/C_{0}$ become equivalent when $l\to\infty$ because the variance of their difference vanishes in this limit. Thus, also all higher statistical moments of $C_{l}/C_0$ must approach the ones of $C_{\infty}/C_0$ when $l \gg 1$. This convergence can be intuitively understood by noting that $P_{l}(\hat x_i\cdot \hat x_j)$ is a rapidly fluctuating function on the sphere when $l\gg 1$ and thus the contributions from different binaries in the second term tend to cancel each other. 

As can be seen from the distributions in Fig.~\ref{fig:PdfClC0}, the distributions for $C_{l\leq 3}/C_0$ are noticeably wider than the distribution of $C_{\infty}/C_0$. This is in agreement with Eq.~\eqref{eq:sigma}. A good agreement with the distribution of $C_{\infty}/C_0$ is found at higher multipoles, when $l \geq 10$.

\subsection{Weak/strong source split} 

The high-$\Omega$ tail of the distribution~\eqref{eq:dNdOmega} means that a handful of strong sources typically dominate the signal even when the total number of sources can be several orders of magnitude larger. According to Fig.~\ref{fig:bestfit}, half the signal in any of the NG15 frequency bins is generated by at most a few hundred contributions in all models considered here, while the total number of sources can be up to order $10^5$. The strongest sources will also dominate the anisotropy. 

Following~\cite{Ellis:2023dgf, Raidal:2024odr}, we model the weak and strong sources separately. At a given frequency, a source is considered strong, if its GW signal exceeds some fixed threshold $\Omega \geq \Omega_{\rm thr}$, which we will specify later. The total luminosity can then be decomposed as
\bea
    \Omega &= \Omega_S + \Omega_W\,.
\eea
The strong and weak components, $\Omega_S$ and $\Omega_W$, consist of the individual contribution from all the strong and weak binaries\footnote{Note that $\Omega_i$ differ from $\Omega$ in $P^{(1)}(\Omega|f)$ by a factor of $\ln(f_{i+1}/f{i})$, that is, by the logarithmic width of the bin\cite{Ellis:2023dgf}.} 
\be
    \Omega_S \equiv \sum_{i \in I_W} \Omega_{i}\,,\qquad
    \Omega_W \equiv \sum_{i \in I_S} \Omega_{i}\,,
\ee
where $I_W$, $I_S$ denote the index sets of weak and strong sources, respectively.
The multipole ratio~\eqref{clc0} then takes the form 
\bea\label{eq:clc0_split_0}
    \frac{C_{l}}{C_0} 
    &= \sum_{i,j\in  I_W} r_{i} r_{j} P_{l}(\hat x_{i}\cdot \hat x_{i}) \\
    &+ 2\!\!\sum_{i \in  I_W,j\in  I_S} r_{i} r_{j} P_{l}(\hat x_{i}\cdot \hat x_{j}) \\
    &+ \sum_{i,j \in  I_S} r_{i} r_{j} P_{l}(\hat x_{i}\cdot \hat x_{j}) \,,
\eea
Averaging over the strengths and sky locations of the weak sources and expanding in the variance of the weak source luminosity gives (see Appendix~\ref{app:stat})
\bea\label{eq:clc0_approx}
    \frac{C_{l}}{C_0} 
    &\approx \frac{\bar{\Omega}_W^2/N_W}{(\Omega_S + \bar{\Omega}_W)^2 } 
    + \sum_{i,j \in  I_S} \bar{r}_{i} \bar{r}_{j} P_{l}(\hat x_{i_S}\cdot \hat x_{j_S}) \,,
\eea
where $\bar{\Omega}_W \equiv \langle\Omega_W\rangle$, $N_W$ is the number of weak sources and we defined the rescaled ratio
\be\label{eq:bar_ri}
    \bar{r}_i \equiv \frac{\Omega_i}{\Omega_S + \bar{\Omega}_W} 
\ee
This approximation gives the leading contribution when expanding in $\sigma^2_{\Omega_W}/(\Omega_S + \Omega_W)^2$, that is, it effectively ignores fluctuations in the luminosity of weak sources. We checked that including the $\sigma^2_{\Omega_W}$ correction gives at most a 1\% correction to our estimates. In Eq.~\eqref{eq:clc0_approx}, the individual luminosities of strong sources and thus also their total contribution $\Omega_S$ is random and will be modelled using a MC approach. As long as $\Omega_S \gtrsim \bar{\Omega}_W\sqrt{N_S/N_W}$, the dominant contribution arises from the last term in Eq.~\eqref{eq:clc0_approx}.

\section{Results}
\label{sec:results}

\begin{figure}
    \centering    
    \includegraphics[width=0.97\columnwidth]{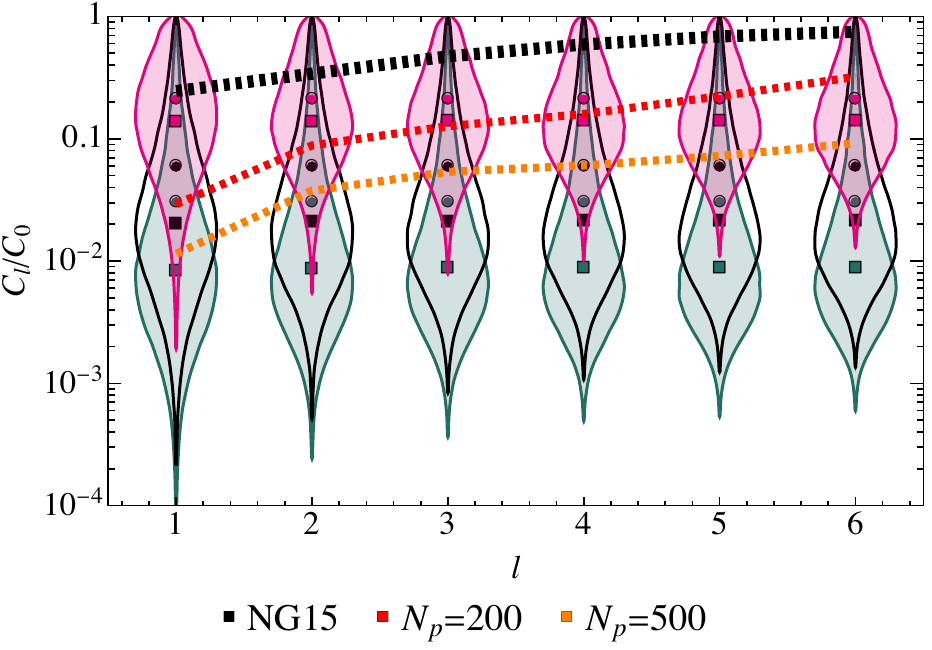}
    \vspace{-2mm}
    \caption{The ratio between the zeroth order and higher-order coefficients of the GW anisotropy power spectrum. The distributions shown are obtained from $10^5$ sky realizations at a GW frequency of $1.98$\,nHz. The medians and means of each distribution are shown by squares and circles respectively. The black dashed curve shows the NG15 upper bound~\cite{NANOGrav:2023tcn} and the other dashed curves the future prospects assuming an EPTA-like noise and a 15-year observation time~\cite{Depta:2024ykq}. The colours of the violins represent the same models as in Fig.~\ref{fig:bestfit}.}
    \label{fig:anisotropies}
\end{figure}

We choose the threshold strength of a strong source, $\Omega_{\rm thr}$, such that the number of strong sources is $N_S=50$,\footnote{Note that, although $N_{0.5}>50$ for the eccentric model at low frequencies, the source population is very isotropic and the number of strongest sources which contribute most to the anisotropy, is still small.} generate $10^5$ realizations of the population of strong sources at a given frequency and compute the distributions of the normalised power spectrum coefficients as described in Sec.~\ref{sec:models}.
The resulting distributions are shown in Fig.~\ref{fig:anisotropies}. The mean value for the power spectrum coefficients remains constant for $\ell\geq1$, with the value given by~\eqref{eq:ClMean}. However, there are differences between the distributions of the coefficients not only between models but also between multipoles. As was seen above, the distribution is broadest at the first multipole and, at higher multipoles, it quickly becomes almost $l$ independent.

\begin{figure}
    \centering    
    \includegraphics[width=0.97\columnwidth]{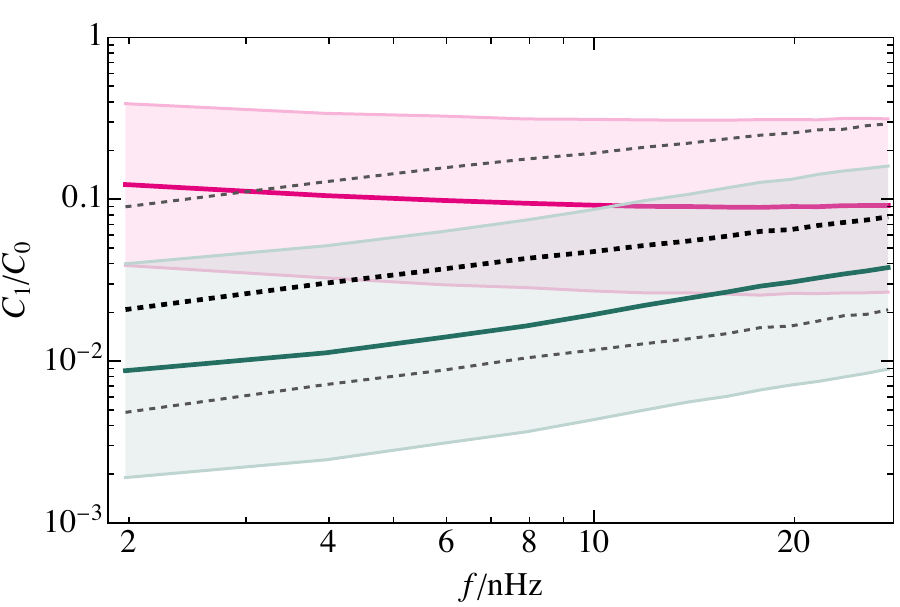}
    \vspace{-2mm}
    \caption{Frequency scaling of the median normalised power spectrum coefficient at $l=1$ for each of the three models considered. The shaded areas show the $1\sigma$ confidence levels around the median. The colours represent the same models as in Fig.~\ref{fig:bestfit}. The GW-only model is shown in dashed for better visibility.}
    \label{fig:freq}
\end{figure}

The magnitude of the anisotropy can partly be understood by looking at the number of sources contributing to the GW background for each model, shown in the left panel of Fig.~\ref{fig:bestfit}. Environmental effects directly reduce the number of sources in each PTA frequency band, leading to the smallest number of sources and hence the largest anisotropy. Emission across a range of frequencies due to an eccentric orbit results in each binary contributing to multiple frequency bins (as well as binaries with primary emission peaks outside the PTA range contributing to the signal). This results in the largest number of sources and the smallest anisotropy.

The anisotropy of the GW background also depends on the frequency of the measured GWs. This is a combined effect of the change in the number of sources and the strength of GW emission. As shown in~\cite{Sato-Polito:2023spo}, the mean anisotropy increases with frequency according to a simple power law at low frequencies with a model-dependent flattening at higher frequencies. Additional environmental interactions modify the power law and can, for some parameter values, lead to a saddle-shaped curve (see Fig. 4 of~\cite{Sato-Polito:2023spo}). Our results broadly agree with this behaviour, as can be seen in Fig~\ref{fig:freq} which shows the intervals containing $68\%$ of the realizations in each frequency band for the three models considered here. Moreover, Fig.~\ref{fig:freq} shows a distinct separation between the best-fit model anisotropies at the lowest frequency bands. Although the $68\%$ bands of the eccentric and environmental models for the first multipole moment slightly overlap, these bands separate fully at higher multipoles due to the narrowing of their distributions.

\section{Discussion}
\label{sec:discussion}

\begin{figure*}
    \centering    
    \includegraphics[width=0.97\textwidth]{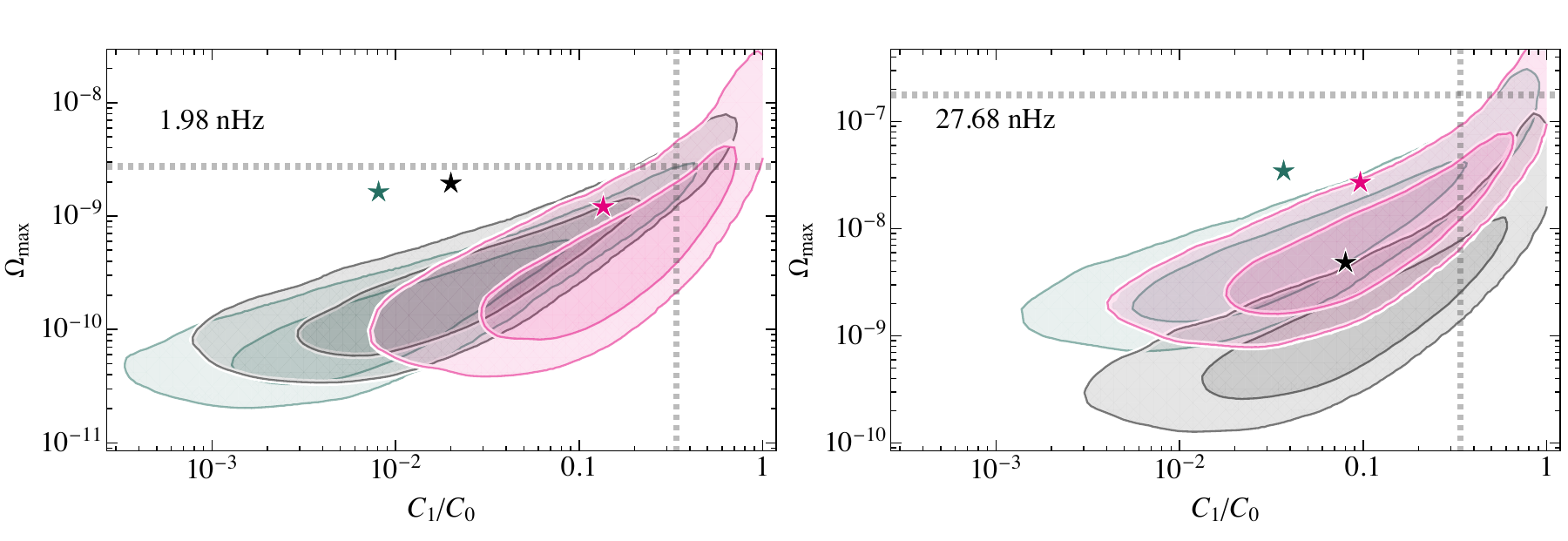}
    \caption{The GW contribution from the loudest binary versus the first multipole at the $f = 1.98\,$nHz (left) and $f = 27.68\,$nHz (right) frequency bins. The darker and lighter shaded contours show regions containing 68\% and 95\% of the realizations, respectively, and the colours represent the same models as in Fig.~\ref{fig:bestfit}. The star denotes the median of $(\Omega_{\rm tot}, C_1/C_0)$. The grey dashed vertical and horizontal lines denote the current NG15 upper bound for $C_1/C_0$~\cite{NANOGrav:2023tcn} and individual SMBH binary detection~\cite{NANOGrav:2023pdq}, respectively.}
    \label{fig:correlation}
\end{figure*}

\subsection{Prospects for future observations}

As can be seen from Fig.~\ref{fig:anisotropies}, the current NG15 upper bounds on measured anisotropy are sufficiently strong to start constraining environmentally-driven SMBH models, with the mean value of the $C_1/C_0$ distribution being comparable to the current NG15 bound. However, as the distribution of $C_1/C_0$ is long-tailed, the most likely value of the distribution remains below the current upper bound. As PTAs add pulsars to their catalogues, these upper bounds on anisotropy will lower. Possible future bounds for EPTA-like noise curves and 15-year observational times derived in~\cite{Depta:2024ykq} are also shown in Fig.~\ref{fig:anisotropies} for different numbers of pulsars observed. For $N_p=200$, the environmental model discussed in this paper can be almost fully excluded in the case of non-measurement. The anisotropy of circular GW emission only and eccentric SMBH binary models are likely to be probed only at much larger numbers of pulsars, $N_p\geq 500$. However, the prospects improve also with an increase in the precision of the timing measurements. As shown in~\cite{Depta:2024ykq}, the prospective upper bounds for $C_l/C_0$ with SKA-like noise for the same number of pulsars are almost an order of magnitude stronger than with EPTA-like noise.

\subsection{Spectral fluctuations and resolvable sources}

GW backgrounds dominated by a few sources tend to be more anisotropic than the ones dominated by many sources. This results in a positive correlation between anisotropy and spectral fluctuations. A naive quantitative estimate can be made by noting that, with $N$ GW sources, the average over the sky locations satisfies $\langle C_l/C_0 \rangle_{\hat x} = C_\infty/C_0 \geq 1/N$, and that this inequality is saturated when all the sources are equally loud, $r_i = 1/N$. Without the sky average, downward fluctuations are possible due to cancellations between the Legendre polynomials in~\eqref{clc0}. It is possible to further refine this inequality and to find the lower and upper bound
\be\label{eq:C_l_bounds}
    r_{\rm max} \geq \frac{C_\infty}{C_0} \geq \frac{p^2}{N_{p}} \,,
\ee
where $N_{p}$ is the number of loudest sources contributing the fraction $p$ to the signal and $r_{\rm max}$ is the fractional strength of the loudest signal. The upper bound follows from~\eqref{eq:C_infty} after noting that the anisotropy is maximized if the signal consists of $N = 1/r_{\rm max}$ sources with maximal strength $r_{\rm max}$. When considering only the strongest source, the lower bound reads $C_\infty/C_0 \geq r_{\rm max}^2$ since $N_{p} = 1$ and $p = r_{\rm max}$.

The lower bound quantifies the intuition that the anisotropy is large when the signal is dominated by a few strong sources. The upper bound says that a large anisotropy implies the presence of sources that contribute a significant fraction to the signal. Importantly, Fig.~\ref{fig:PdfClC0} shows a nonvanishing probability that $C_\infty/C_0$ is practically one in which case a single source must completely dominate the GW signal. 

To account for the sky locations of the SMBH binaries in $C_l/C_0$ in specific realizations of the SMBH binary population, we need to resort to numerical methods. In Fig.~\ref{fig:correlation}, we show the distributions between the strongest \emph{individual} GW source and the dipole, $C_1/C_0$, obtained from $10^4$ realizations of the SMBH population, at both $f=1.98 \rm nHz$ and $f=27.68 \rm nHz$. The darker and lighter shaded contours contain 68\% and 95\% of the realizations, respectively. There is a clear trend showing that SMBH populations with the loudest binaries tend to be more anisotropic. When $\Omega_{\rm max}\approx \Omega_{\rm tot}$ or $r_{\rm max} \approx 1$, we find that $C_l/C_0 \approx (\Omega_{\rm max}/\Omega_{\rm tot})^2\equiv r_{\rm max}^2$ suggesting that the anisotropy is dominated by a single source in these realizations. 

The median realizations of the \emph{total} GW signal and $C_1/C_0$ are shown by stars in Fig.~\ref{fig:correlation} (they match the squares in Fig.~\ref{fig:anisotropies}). Thus, $\Omega_{\rm max}$ values comparable to or higher than the star are likely to correspond to realizations in which the strongest binary makes up a sizeable fraction of the GW signal, so it would also be responsible for an upward spectral fluctuation at that frequency. This is unlikely unless the stars are within the corresponding shaded areas, that is, at higher frequencies or for models involving environmental effects.

The dependence of these correlations on GW frequency is non-trivial. However, the general characteristics of the correlations reflect the behaviour seen in Figs.~\ref{fig:freq} and~\ref{fig:bestfit}. For the environmental model, the number of loudest binaries remains constant whilst it decreases for the eccentric and GW-only models. The total amplitude of the GW signal increases with frequency for all models. Correspondingly, the loudest contribution, $\Omega_{\rm max}$ increases in both magnitude and strength relative to the other sources for all models. This shifts the contours higher and leads to an increase in anisotropy, seen as a shift to higher values of $C_1/C_0$ in the contours. Note that, for the GW-only model the increase in spectral amplitude is small, making the shifts in $\Omega_{\rm max}$ and $C_1/C_0$ much smaller when compared with the other models.

The dashed lines in Fig.~\ref{fig:correlation} show the NG15 constraints on $C_1/C_0$~\cite{NANOGrav:2023tcn} and on the individual detection of a SMBH binary~\cite{NANOGrav:2023pdq}. Currently, the non-observation of both of these quantities mostly excludes the high-amplitude end of spectral and anisotropy fluctuations.

\subsection{Including cosmological backgrounds}

It is possible that the PTA GW background does not arise purely from SMBH mergers but contains also a cosmological component $\Omega_{\rm CGWB}(f)$ originating from the primordial Universe. Many cosmological models can fit the recent PTA data equally well or even mildly better than the SMBH models~\cite{Ellis:2023oxs}.  In such cases, the GW spectrum is decomposed as
\be
    \Omega(f) = \Omega_{\rm SMBH}(f) + \Omega_{\rm CGWB}(f)\,.
\ee
In contrast to the GW background from SMBHs, $\Omega_{\rm SMBH}(f)$, we expect $\Omega_{\rm SMBH}(f)$ to be isotropic and smooth, that is, without notable spectral fluctuations. 

To accommodate such a background into the formalism above, the relative contribution from a single binary must be rescaled (as above, we'll drop the explicit frequency dependence)
\be\label{eq:r_CGWB}
    r_i 
    = \frac{\Omega_i}{\Omega_{\rm SMBH} + \Omega_{\rm CGWB}}
    \equiv r_{i,\rm BH}\, r_{\rm SMBH}\,
\ee
where $r_{i,\rm BH} \equiv \Omega_i/\Omega_{\rm SMBH}$ is the relative strength from an individual binary compared to the GW signal from all binaries, and $r_{\rm SMBH} \equiv \Omega_{\rm SMBH}/(\Omega_{\rm SMBH} + \Omega_{\rm CGWB})$ is the fraction of the total signal from binaries. The anisotropy for a particular realization of the SMBH population can then be described by Eq.~\eqref{clc0}.

Homogeneous cosmological backgrounds with negligible statistical fluctuations can easily included in the approach summarized by Eq.~\eqref{eq:clc0_approx} in which the binary sources are split into weak and strong ones. To include the cosmological background the shift
\be
    \bar{\Omega}_W \to \bar{\Omega}_W + \Omega_{\rm CGWB}
\ee
should be performed \emph{only} in the denominator Eqs.~\eqref{eq:clc0_approx} and~\eqref{eq:bar_ri}. The numerator in Eq.~\eqref{eq:clc0_approx} gives the contribution to the variance from weak binaries and is not affected by the cosmological background. Nevertheless, Eq.~\eqref{eq:clc0_approx} demonstrates that there is a degeneracy between the strength of cosmological GW backgrounds $\Omega_{\rm CGWB}$ and the leading characteristics $\bar{\Omega}_W$, $N_W$ of the weak SMBH binary populations.

A rough estimate of the contribution from a cosmological GW background can be obtained by replacing the last term in~\eqref{eq:r_CGWB} by its average, $r_i \approx r_{i,\rm BH} \langle r_{\rm SMBH} \rangle$. In that case, we find that compared to the SMBH-only case, the anisotropy is reduced by a factor of $\langle r_{\rm SMBH} \rangle^2$, that is
\be\label{eq:Cl_scaling}
    \frac{C_{l\geq1}}{C_0} \approx \left.\frac{C_{l\geq1}}{C_0}\right|_{\Omega_{\rm CGWB} = 0} \langle r_{\rm SMBH} \rangle^2\,.
\ee
Statistically, this approximation amounts to ignoring correlations between $r_{i,\rm BH}$ and $r_{\rm SMBH}$. However, as $r_{\rm SMBH}$ tends to have a heavy tail created mainly due to the loudest SMBH, that is, the largest $r_{i,\rm BH}$, we expect that such correlations can have a sizeable $\mathcal{O}(1)$ effect on the $C_l/C_0$ and so the approximate scaling~\eqref{eq:Cl_scaling} can provide an order of magnitude estimate at best.

A mixed signal from SMBHs and cosmological sources would be harder to probe using anisotropies due to uncertainty in the proportion of signal coming from SMBHs and difficulty in resolving lower values of the anisotropy power spectrum coefficients.

\section{Conclusions}
\label{sec:concl}

We have studied the statistical properties of the anisotropy of GW backgrounds from SMBHs. We have derived the distributions of the (normalised) anisotropy power spectrum coefficients, $C_l/C_0$, and characterised their dependence on multipole number and frequency. We have shown that these distributions are widest at $l=1$, and, assuming statistical isotropy, tend towards an $l$-independent distribution that is characterized fully by the distribution of GW luminosities of the SMBH binaries.

We have investigated the anisotropy distributions of three models of SMBH binary evolution in benchmark cases found to best fit the GW background spectrum inferred from the NANOGrav 15-year measurements. The model incorporating binary eccentricities predicts smaller anisotropies than the model where binaries are circular but are affected by environmental interactions because, in the latter case, the GW signal is typically dominated by less than 10 sources. 

We have compared the predicted anisotropies with the current upper bounds finding that the model incorporating environmental effects is in a mild tension with the bounds. We have also made comparisons to the projected upper bounds for future measurements with expanded pulsar catalogues. A catalogue of $N_p=200$ pulsars and a 15-year measurement time would be sufficient to nearly fully exclude the environmental effect best fit and would also probe the anisotropies in the model with circular binaries evolving purely by GW emission. The sensitivity needed to investigate the anisotropy generated by eccentric binaries will only be reached with experiments with $N_p\geq500$. These prospects hold under the assumption that the GW background originates purely from SMBH binaries. For signals of mixed cosmological and astrophysical origin, the anisotropy is lowered.

Large upward fluctuations in the GW amplitude are typically accompanied by a heightened anisotropy. We have quantified the relation between anisotropy and the GW luminosity of the strongest source. As this relation varies from model to model, determining the correlation between spectral fluctuations and anisotropy opens new possibilities for SMBH model discrimination.

\begin{acknowledgments}
This work was supported by the Estonian Research Council grants PRG803, PSG869, RVTT3 and RVTT7 and the Center of Excellence program TK202. The work of V.V. was partially funded by the European Union's Horizon Europe research and innovation program under the Marie Sk\l{}odowska-Curie grant agreement No. 101065736.
\end{acknowledgments}

\appendix
\section{Modelling eccentricity and environmental effects}
\label{app:ecc}

A general eccentric binary emits GWs in several harmonic modes with the power emitted in the $n$-th harmonic being~\cite{Peters:1963ux}
\bea \label{eq:En}
    \frac{\td E_n}{\td t_r}&=\frac{32}{5}(2\pi\mathcal{M}f_{\rm b})^{10/3}g_n(e)
    \equiv \frac{\td E_c}{\td t_r}g_n(e) \,,
\eea
where $\mathcal{M}$ denotes the chirp mass, $f_{\rm b}$ the orbital frequency and $z$ the redshift of the binary, and $\td E_c/\td t_r$ is the power emitted by a circular binary with the above binary parameters. The observer frame time $t$ is related to the time $t_r$ in the rest frame of the binary by $\td t = (1+z) \td t_r$. The relative power radiated in the $n$-th harmonic is
\bea \label{eq:gn}
    &g_n(e) = \frac{n^4}{32} \Bigg[ \frac{4}{3n^2}J_n^2(ne) + \bigg(J_{n-2}(ne)-2eJ_{n-1}(ne) \\
    &\hspace{8pt} + \frac{2}{n}J_n(ne)+2eJ_{n+1}(ne)-J_{n+2}(ne)\bigg)^{\!2} \\
    &\hspace{8pt} + (1-e^2)\bigg(J_{n-2}(ne)-2J_n(ne)+J_{n+2}(ne)\bigg)^{\!2} \,\Bigg] ,
\eea
where $J_n$ are the $n$-th Bessel functions of the first kind. Note that, for circular binaries, $g_n(0)=\delta_{2,n}$.

The total power emitted by a binary across all frequencies is
\be\label{eq:Etot}
     \frac{\td E_{\rm GW}}{\td t_r} = \frac{\td E_c}{\td t_r}\mathcal{F}(e) \,,
\ee
where
\be
     \mathcal{F}(e) = \sum_{n=1}^\infty g_n(e)=\frac{1+73/24e^2+37/96e^4}{(1-e^2)^{7/2}} \,.
\ee
Note that $\mathcal{F}(0)=1$.

The orbital dynamics of a black hole binary is given by the evolution of its orbital frequency and eccentricity~\cite{Peters:1963ux}
\bea \label{eq:orbital}
    \frac{\td \ln f_{\rm b}}{\td t_r} 
    &= \frac{96}{5} \mathcal{M}^{5/3} (2\pi f_{\rm b})^{8/3} \mathcal{F}(e)  \,, \\
    \frac{\td e}{\td t_r} 
    &= -\frac{1}{15} \mathcal{M}^{5/3} (2\pi f_{\rm b})^{8/3} \mathcal{G}(e) \,,
\eea
where
\be \label{eq:Ge}
    \mathcal{G}(e) = \frac{304e+121e^3}{(1-e^2)^{5/2}} \,.
\ee
From this, the observed residence time of a binary at orbital frequency $f_{\rm b}$ is
\bea\label{eq:residence1}
    \frac{\td t}{\td \ln f_{\rm b}} 
    &= \frac{5}{96} \frac{1+z}{\mathcal{M}^{5/3}} (2\pi f_{\rm b})^{-8/3} \frac{1}{\mathcal{F}(e)} 
    \equiv \frac23 t_{\rm GW}
\eea
where $t_{\rm GW} \equiv |E|/\dot E_{\rm GW}$ is the characteristic timescale of energy loss through GW emission.
The factor $1+z$ arises when converting the binary rest frame time $t_r$ to the time $t$ in the observer frame. 

Following~\cite{Raidal:2024odr}, we model the binary eccentricities by fixing the distribution of $e$ at the lowest NG15 frequency bin. We consider a power-law distribution $P(e) = (\gamma+1) e^\gamma$ and trade $\gamma$ for the mean initial eccentricity, $\langle e\rangle_{2\,{\rm nHz}} = (1+\gamma)/(2+\gamma)$, which is the only free parameter of our eccentricity model.

The residence time can be further modified due to interactions between the SMBH binary and its environment. Their exact impact is unknown and, for the sake of simplicity and generality, we introduce them via the environmental timescale $t_{\rm env} \equiv |E|/\dot E_{\rm env}$ of energy loss, for which will adapt the power-law ansatz~\cite{Ellis:2023dgf},
\be
    t_{\rm env} = \left( \frac{2f_b}{f_{\rm GW}} \right)^\alpha \,, \quad f_{\rm GW} = f_{\rm ref} \left( \frac{\mathcal{M}}{10^9 \Msun} \right)^\beta \,,
\ee
and fix $\alpha = 8/3$ and $\beta = 5/8$ so that $f_{\rm ref}$ is the only free parameter of the model. Environmental interactions lead to the shortening of the residence time of a binary as given by
\bea
     \frac{\td t}{\td \ln f_{\rm b}} 
     &= \frac23 t_{\rm GW} \left[1 + \frac{t_{\rm GW}}{t_{\rm env}}\right]^{-1} 
\eea
The orbital frequency evolution of the binary eccentricity is affected in a similar manner~\cite{Raidal:2024odr},
\be\label{eq:eccenvevolve}
    \frac{\td e}{\td \ln f_{\rm b}} 
    = -\frac{1}{288}\frac{\mathcal{G}(e)}{\mathcal{F}(e)}\left[1 + \frac{t_{\rm GW}}{t_{\rm env}}\right]^{-1}\,.
\ee

\section{Statistics of $C_l/C_0$}
\label{app:stat}

Here we list some of the identities used in the derivation of the results in the main text. 

The angular averages involve only the Legendre polynomials. Most of the results stem from the identity
\be
    \int \frac{\td \Omega_i}{4\pi} P_l(\hat x_m \cdot \hat x_i)P_{l'}(\hat x_i \cdot \hat x_n)
    = \frac{\delta_{ll'}}{2l + 1} P_l(\hat x_m \cdot \hat x_m)
\ee
For $l>0$ we have that
\bea
    \langle P_l(\hat x_i \cdot \hat x_j) \rangle_{\hat x_i} 
    &= 0, \\
    \langle P_l(\hat x_i \cdot \hat x_j) P_l(\hat x_i \cdot \hat x_m) \rangle 
    &= \delta_{ij}\delta_{mn} \\
    &+ \frac{\delta_{im}\delta_{jn}+\delta_{in}\delta_{jm} - 2\delta_{ij}\delta_{mn}}{2l +1},, \nonumber
\eea
where the first average is taken only over $\hat x_i$.

To compute higher moments, we will split $C_{l}/C_0$ into two pieces
\be
    \frac{C_{l}}{C_0} 
    = \sum_i r_i^2 + \sum_{i\neq j} r_i r_j P_l(\hat x_i \cdot \hat x_j) 
    \equiv \frac{C_{\infty}}{C_0}  + \Delta_l  
\ee
which can be used to study specific statistical properties of $C_{l}/C_0$. Importantly, $C_{\infty}/C_0$ does not depend on the spatial distribution. It is fully determined by the distribution of the signal's relative loudness. The following expectation values can be estimated
\bea\label{eq:averages}
    \langle \Delta_l \rangle 
    &= \left\langle \Delta_l \frac{C_{\infty}}{C_0} \right\rangle = 0 \,,
    \\
    \langle \Delta_l^2 \rangle 
    &= \sum_{i\neq j,m\neq n} \langle r_i r_j r_m r_n \rangle \langle P_l(\hat x_i \cdot \hat x_j) P_l(\hat x_m \cdot \hat x_n) \rangle \,,
    \\
    &= \frac{2}{2l+1}\sum_{i\neq j} \langle r_i^2 r_j^2\rangle \,,
    \\
    \left\langle \frac{C_{\infty}}{C_0} \right\rangle
    &= \sum_i \left\langle r_i^2 \right\rangle \,,
    \\
    \left\langle \left(\frac{C_{\infty}}{C_0} \right)^2\right\rangle 
    &= \sum_{i} \left\langle r_i^4\right\rangle + \sum_{i\neq j} \left\langle r_i^2 r_j^2 \right\rangle\,,
\eea
and thus
\bea
    \left\langle C_{l}/C_0 \right\rangle 
    &= \left\langle C_{\infty}/C_0 \right\rangle\,,
    \\
    \sigma_{C_{l}/C_0}^2
    &= \sigma_{C_{\infty}/C_0}^2 + \sigma_{\Delta_l}^2\,
\eea
confirming the expressions in Sec.~\ref{sec:anisotropy}.

Since all $r_i$ are statistically identical but not independent, then it's sufficient to the expectation values related to a single or a few sources, such as,
$
    \langle  r_1^m \rangle\,, 
    \langle  r_1^m r_2^n \rangle\,, \ldots,
$
using which we can eliminate the sums in Eq.~\eqref{eq:averages}:
\bea
    \langle \Delta_l^2 \rangle 
    &= \frac{2 N_{\rm b} (N_{\rm b}-1)}{2l+1}  \langle r_1^2 r_2^2\rangle \,,
    \\
    \left\langle \frac{C_{\infty}}{C_0} \right\rangle
    &= N_{\rm b} \left\langle r_1^2 \right\rangle\,,
    \\
    \left\langle \left(\frac{C_{\infty}}{C_0} \right)^2\right\rangle 
    &= N_{\rm b} \left\langle r_1^4\right\rangle + N_{\rm b}(N_{\rm b}-1)\left\langle r_1^2 r_2^2 \right\rangle\,.
\eea

Let us now consider the average over the weak sources of Eq.~\eqref{eq:clc0_split_0}, which we express here as
\be
    C_{l}/C_0 = I_{W} + I_{WS} + I_{S} \,,\nonumber
\ee
where
\bea
    \mathcal{I}_{W} & \equiv  \sum_{i,j\in  I_W} r_{i} r_{j} P_{l}(\hat x_{i}\cdot \hat x_{i}) \\
    \mathcal{I}_{WS} &\equiv 2\!\!\sum_{i \in  I_W,j\in  I_S} r_{i} r_{j} P_{l}(\hat x_{i}\cdot \hat x_{j}) \\
    \mathcal{I}_{S} &\equiv \sum_{i,j \in  I_S} r_{i} r_{j} P_{l}(\hat x_{i}\cdot \hat x_{j}) \,
\eea
The mean of the central term vanishes, $\left\langle \mathcal{I}_{WS}\right\rangle_W = 0$ because the positions of the weak sources are not correlated with the strong ones (the subindex $W$ denotes averaging over weak sources only).

The only relevant terms are thus $\mathcal{I}_{S}$ and $\mathcal{I}_{W}$ which we'll compute up to the order $\sigma_{\Omega_W}^2$. To simplify the derivation, we'll denote
\bea
    \delta \Omega_i &\equiv \Omega_i - \bar\Omega_1\,, \qquad
    \Omega_{W,i} \equiv \Omega_{W} - \delta \Omega_i\,,
\eea
where $\bar\Omega_1 \equiv \langle \Omega_i \rangle_W$ is the average of a single source. $\Omega_{W,i}$ is the total strength of weak sources with the $i$-th source replaced by its average. Since $\Omega_i$ are independent, then so are $\Omega_{W,i}$ and $\delta \Omega_i$. We will consider an expansion in $\delta \Omega_i$ as it is expected to be small compared to both $\Omega_S$ and $\Omega_W$. 

\begin{widetext}
Moreover, $\mathcal{I}_{W}$ is simplified because $\langle P_{l}(\hat x_{i}\cdot \hat x_{j}) \rangle_W = \delta_{ij}$ so that
\bea
    \left\langle\mathcal{I}_{W} \right\rangle_{W}
    &= \sum_{i\in  I_W} \left\langle r_{i}^2 \right\rangle_{W}
    = \sum_{i\in  I_W} \left\langle \frac{\Omega_i^2}{(\Omega_S + \Omega_{W})^2} \right\rangle_{W} 
    = \sum_{i\in  I_W} \left\langle \frac{(\bar\Omega_1 + \delta\Omega_i)^2}{(\Omega_S + \Omega_{W,i} + \delta\Omega_i)^2} \right\rangle_{W} \\
    &= \sum_{i\in  I_W} \left\langle 
    \frac{\bar\Omega_1^2 + \langle\delta\Omega_i^2\rangle}{(\Omega_S + \Omega_{W,i})^2} 
    - \frac{4 \bar\Omega_1 \langle\delta\Omega_i^2\rangle}{(\Omega_S + \Omega_{W,i})^3}
    + \frac{3\bar\Omega_1^2 \langle\delta\Omega_i^2\rangle}{(\Omega_S + \Omega_{W,i})^4}
    + \mathcal{O}(\delta\Omega_i^3)
    \right\rangle_{W}\\
    &= 
    \frac{\bar\Omega_W^2 N_W^{-1}}{(\Omega_S + \bar\Omega_W)^2} 
    +
    \frac{\sigma^2_{\Omega_W}}{(\Omega_S + \bar\Omega_W)^2} \left[1
    - \frac{4 \bar\Omega_W N_W^{-1}}{\Omega_S + \bar\Omega_W}
    + \frac{3 \bar\Omega_W^2 N_W^{-1}}{(\Omega_S + \bar\Omega_W)^2} \right]
    + \mathcal{O}(\sigma_{\Omega_W}^3)
    \,,
\eea
where we dropped terms linear in $\delta\Omega_i$ as their averages vanish. We also use $\bar\Omega_1 = \bar\Omega_W/N$ and $\langle\delta\Omega_i^2\rangle = \sigma^2_{\Omega_W}/N_W$, where $N_W$ is the number of weak sources. In the last step, we estimated the average of the denominator
\be\label{eq:E_frac_W}
    \left\langle \frac{1}{(\Omega_S + \Omega_{W})^n} \right\rangle_{W}
    =   \frac{1}{(\Omega_S + \bar\Omega_{W})^n}
    + \frac{n (n+1)}{2}\frac{\sigma^2_{\Omega_W}}{(\Omega_S + \bar\Omega_{W})^{n+2}} + \mathcal{O}(\sigma^4_{\Omega_W})
\ee
by expanding in $\delta\Omega_{W} \equiv \Omega_{W} - \bar\Omega_{W}$. Note that $\langle\Omega_{W,i}\rangle = \langle\Omega_{W}\rangle \equiv \bar\Omega_{W}$ and $\sigma^2_{\Omega_{W,i}} = (1-N_W^{-1})\sigma^4_{\Omega_W}$.

A similar procedure gives
\bea     
    \left\langle \mathcal{I}_{S} \right\rangle_{W} 
    &= \left\langle \frac{1}{(\Omega_S + \Omega_{W})^2} \right\rangle_{W} \sum_{i,j \in  I_S} \Omega_{i} \Omega_{j} P_{l}(\hat x_{i}\cdot \hat x_{j})\, 
\eea
where average is given by~\eqref{eq:E_frac_W}.
\end{widetext}
The approximation given in Eq.~\eqref{eq:clc0_split_0} for the weak-strong split is obtained by dropping the $\mathcal{O}(\sigma^2_{\Omega_W})$ terms. Including the $\mathcal{O}(\sigma^2_{\Omega_W})$ correction affects the results by less than 1\% with $N_S = 50$ strong sources. The corrections would be further suppressed when $\Omega_W \ll \Omega_S$, which we do not require. 

We also note that we have estimated the corrections when averaging over weak sources due to spectral fluctuations caused by weak sources. However, the averaging procedure itself is an approximation and we expect that the corrections to the width of the distribution of $C_l/C_0$ are of the order $\sigma^2_{\Omega_W}$. The heavy non-Gaussian tail of the $\Omega$ distribution guarantees that such corrections are negligible, that is, the tail can be fully modelled by the strong sources.

\bibliography{refs}

\begin{thebibliography}{42}%
\makeatletter
\providecommand \@ifxundefined [1]{%
 \@ifx{#1\undefined}
}%
\providecommand \@ifnum [1]{%
 \ifnum #1\expandafter \@firstoftwo
 \else \expandafter \@secondoftwo
 \fi
}%
\providecommand \@ifx [1]{%
 \ifx #1\expandafter \@firstoftwo
 \else \expandafter \@secondoftwo
 \fi
}%
\providecommand \natexlab [1]{#1}%
\providecommand \enquote  [1]{``#1''}%
\providecommand \bibnamefont  [1]{#1}%
\providecommand \bibfnamefont [1]{#1}%
\providecommand \citenamefont [1]{#1}%
\providecommand \href@noop [0]{\@secondoftwo}%
\providecommand \href [0]{\begingroup \@sanitize@url \@href}%
\providecommand \@href[1]{\@@startlink{#1}\@@href}%
\providecommand \@@href[1]{\endgroup#1\@@endlink}%
\providecommand \@sanitize@url [0]{\catcode `\\12\catcode `\$12\catcode `\&12\catcode `\#12\catcode `\^12\catcode `\_12\catcode `\%12\relax}%
\providecommand \@@startlink[1]{}%
\providecommand \@@endlink[0]{}%
\providecommand \url  [0]{\begingroup\@sanitize@url \@url }%
\providecommand \@url [1]{\endgroup\@href {#1}{\urlprefix }}%
\providecommand \urlprefix  [0]{URL }%
\providecommand \Eprint [0]{\href }%
\providecommand \doibase [0]{http://dx.doi.org/}%
\providecommand \selectlanguage [0]{\@gobble}%
\providecommand \bibinfo  [0]{\@secondoftwo}%
\providecommand \bibfield  [0]{\@secondoftwo}%
\providecommand \translation [1]{[#1]}%
\providecommand \BibitemOpen [0]{}%
\providecommand \bibitemStop [0]{}%
\providecommand \bibitemNoStop [0]{.\EOS\space}%
\providecommand \EOS [0]{\spacefactor3000\relax}%
\providecommand \BibitemShut  [1]{\csname bibitem#1\endcsname}%
\let\auto@bib@innerbib\@empty
\bibitem [{\citenamefont {Agazie}\ \emph {et~al.}(2023{\natexlab{a}})\citenamefont {Agazie} \emph {et~al.}}]{NANOGrav:2023gor}%
  \BibitemOpen
  \bibfield  {author} {\bibinfo {author} {\bibfnamefont {G.}~\bibnamefont {Agazie}} \emph {et~al.} (\bibinfo {collaboration} {NANOGrav}),\ }\href {\doibase 10.3847/2041-8213/acdac6} {\bibfield  {journal} {\bibinfo  {journal} {Astrophys. J. Lett.}\ }\textbf {\bibinfo {volume} {951}},\ \bibinfo {pages} {L8} (\bibinfo {year} {2023}{\natexlab{a}})},\ \Eprint {http://arxiv.org/abs/2306.16213} {arXiv:2306.16213 [astro-ph.HE]} \BibitemShut {NoStop}%
\bibitem [{\citenamefont {Antoniadis}\ \emph {et~al.}(2023{\natexlab{a}})\citenamefont {Antoniadis} \emph {et~al.}}]{EPTA:2023fyk}%
  \BibitemOpen
  \bibfield  {author} {\bibinfo {author} {\bibfnamefont {J.}~\bibnamefont {Antoniadis}} \emph {et~al.} (\bibinfo {collaboration} {EPTA, InPTA:}),\ }\href {\doibase 10.1051/0004-6361/202346844} {\bibfield  {journal} {\bibinfo  {journal} {Astron. Astrophys.}\ }\textbf {\bibinfo {volume} {678}},\ \bibinfo {pages} {A50} (\bibinfo {year} {2023}{\natexlab{a}})},\ \Eprint {http://arxiv.org/abs/2306.16214} {arXiv:2306.16214 [astro-ph.HE]} \BibitemShut {NoStop}%
\bibitem [{\citenamefont {Antoniadis}\ \emph {et~al.}(2023{\natexlab{b}})\citenamefont {Antoniadis} \emph {et~al.}}]{EPTA:2023sfo}%
  \BibitemOpen
  \bibfield  {author} {\bibinfo {author} {\bibfnamefont {J.}~\bibnamefont {Antoniadis}} \emph {et~al.} (\bibinfo {collaboration} {EPTA}),\ }\href {\doibase 10.1051/0004-6361/202346841} {\bibfield  {journal} {\bibinfo  {journal} {Astron. Astrophys.}\ }\textbf {\bibinfo {volume} {678}},\ \bibinfo {pages} {A48} (\bibinfo {year} {2023}{\natexlab{b}})},\ \Eprint {http://arxiv.org/abs/2306.16224} {arXiv:2306.16224 [astro-ph.HE]} \BibitemShut {NoStop}%
\bibitem [{\citenamefont {Reardon}\ \emph {et~al.}(2023{\natexlab{a}})\citenamefont {Reardon} \emph {et~al.}}]{Reardon:2023gzh}%
  \BibitemOpen
  \bibfield  {author} {\bibinfo {author} {\bibfnamefont {D.~J.}\ \bibnamefont {Reardon}} \emph {et~al.},\ }\href {\doibase 10.3847/2041-8213/acdd02} {\bibfield  {journal} {\bibinfo  {journal} {Astrophys. J. Lett.}\ }\textbf {\bibinfo {volume} {951}},\ \bibinfo {pages} {L6} (\bibinfo {year} {2023}{\natexlab{a}})},\ \Eprint {http://arxiv.org/abs/2306.16215} {arXiv:2306.16215 [astro-ph.HE]} \BibitemShut {NoStop}%
\bibitem [{\citenamefont {Zic}\ \emph {et~al.}(2023)\citenamefont {Zic} \emph {et~al.}}]{Zic:2023gta}%
  \BibitemOpen
  \bibfield  {author} {\bibinfo {author} {\bibfnamefont {A.}~\bibnamefont {Zic}} \emph {et~al.},\ }\href {\doibase 10.1017/pasa.2023.36} {\bibfield  {journal} {\bibinfo  {journal} {Publ. Astron. Soc. Austral.}\ }\textbf {\bibinfo {volume} {40}},\ \bibinfo {pages} {e049} (\bibinfo {year} {2023})},\ \Eprint {http://arxiv.org/abs/2306.16230} {arXiv:2306.16230 [astro-ph.HE]} \BibitemShut {NoStop}%
\bibitem [{\citenamefont {Reardon}\ \emph {et~al.}(2023{\natexlab{b}})\citenamefont {Reardon} \emph {et~al.}}]{Reardon:2023zen}%
  \BibitemOpen
  \bibfield  {author} {\bibinfo {author} {\bibfnamefont {D.~J.}\ \bibnamefont {Reardon}} \emph {et~al.},\ }\href {\doibase 10.3847/2041-8213/acdd03} {\bibfield  {journal} {\bibinfo  {journal} {Astrophys. J. Lett.}\ }\textbf {\bibinfo {volume} {951}},\ \bibinfo {pages} {L7} (\bibinfo {year} {2023}{\natexlab{b}})},\ \Eprint {http://arxiv.org/abs/2306.16229} {arXiv:2306.16229 [astro-ph.HE]} \BibitemShut {NoStop}%
\bibitem [{\citenamefont {Xu}\ \emph {et~al.}(2023)\citenamefont {Xu} \emph {et~al.}}]{Xu:2023wog}%
  \BibitemOpen
  \bibfield  {author} {\bibinfo {author} {\bibfnamefont {H.}~\bibnamefont {Xu}} \emph {et~al.},\ }\href {\doibase 10.1088/1674-4527/acdfa5} {\bibfield  {journal} {\bibinfo  {journal} {Res. Astron. Astrophys.}\ }\textbf {\bibinfo {volume} {23}},\ \bibinfo {pages} {075024} (\bibinfo {year} {2023})},\ \Eprint {http://arxiv.org/abs/2306.16216} {arXiv:2306.16216 [astro-ph.HE]} \BibitemShut {NoStop}%
\bibitem [{\citenamefont {Hellings}\ and\ \citenamefont {Downs}(1983)}]{Hellings:1983fr}%
  \BibitemOpen
  \bibfield  {author} {\bibinfo {author} {\bibfnamefont {R.~w.}\ \bibnamefont {Hellings}}\ and\ \bibinfo {author} {\bibfnamefont {G.~s.}\ \bibnamefont {Downs}},\ }\href {\doibase 10.1086/183954} {\bibfield  {journal} {\bibinfo  {journal} {Astrophys. J. Lett.}\ }\textbf {\bibinfo {volume} {265}},\ \bibinfo {pages} {L39} (\bibinfo {year} {1983})}\BibitemShut {NoStop}%
\bibitem [{\citenamefont {Antoniadis}\ \emph {et~al.}(2023{\natexlab{c}})\citenamefont {Antoniadis} \emph {et~al.}}]{EPTA:2023xxk}%
  \BibitemOpen
  \bibfield  {author} {\bibinfo {author} {\bibfnamefont {J.}~\bibnamefont {Antoniadis}} \emph {et~al.} (\bibinfo {collaboration} {EPTA}),\ }\href@noop {} {\  (\bibinfo {year} {2023}{\natexlab{c}})},\ \Eprint {http://arxiv.org/abs/2306.16227} {arXiv:2306.16227 [astro-ph.CO]} \BibitemShut {NoStop}%
\bibitem [{\citenamefont {Agazie}\ \emph {et~al.}(2023{\natexlab{b}})\citenamefont {Agazie} \emph {et~al.}}]{NANOGrav:2023hde}%
  \BibitemOpen
  \bibfield  {author} {\bibinfo {author} {\bibfnamefont {G.}~\bibnamefont {Agazie}} \emph {et~al.} (\bibinfo {collaboration} {NANOGrav}),\ }\href {\doibase 10.3847/2041-8213/acda9a} {\bibfield  {journal} {\bibinfo  {journal} {Astrophys. J. Lett.}\ }\textbf {\bibinfo {volume} {951}},\ \bibinfo {pages} {L9} (\bibinfo {year} {2023}{\natexlab{b}})},\ \Eprint {http://arxiv.org/abs/2306.16217} {arXiv:2306.16217 [astro-ph.HE]} \BibitemShut {NoStop}%
\bibitem [{\citenamefont {Ellis}\ \emph {et~al.}(2024{\natexlab{a}})\citenamefont {Ellis}, \citenamefont {Fairbairn}, \citenamefont {H\"utsi}, \citenamefont {Raidal}, \citenamefont {Urrutia}, \citenamefont {Vaskonen},\ and\ \citenamefont {Veerm\"ae}}]{Ellis:2023dgf}%
  \BibitemOpen
  \bibfield  {author} {\bibinfo {author} {\bibfnamefont {J.}~\bibnamefont {Ellis}}, \bibinfo {author} {\bibfnamefont {M.}~\bibnamefont {Fairbairn}}, \bibinfo {author} {\bibfnamefont {G.}~\bibnamefont {H\"utsi}}, \bibinfo {author} {\bibfnamefont {J.}~\bibnamefont {Raidal}}, \bibinfo {author} {\bibfnamefont {J.}~\bibnamefont {Urrutia}}, \bibinfo {author} {\bibfnamefont {V.}~\bibnamefont {Vaskonen}}, \ and\ \bibinfo {author} {\bibfnamefont {H.}~\bibnamefont {Veerm\"ae}},\ }\href {\doibase 10.1103/PhysRevD.109.L021302} {\bibfield  {journal} {\bibinfo  {journal} {Phys. Rev. D}\ }\textbf {\bibinfo {volume} {109}},\ \bibinfo {pages} {L021302} (\bibinfo {year} {2024}{\natexlab{a}})},\ \Eprint {http://arxiv.org/abs/2306.17021} {arXiv:2306.17021 [astro-ph.CO]} \BibitemShut {NoStop}%
\bibitem [{\citenamefont {Bi}\ \emph {et~al.}(2023)\citenamefont {Bi}, \citenamefont {Wu}, \citenamefont {Chen},\ and\ \citenamefont {Huang}}]{Bi:2023tib}%
  \BibitemOpen
  \bibfield  {author} {\bibinfo {author} {\bibfnamefont {Y.-C.}\ \bibnamefont {Bi}}, \bibinfo {author} {\bibfnamefont {Y.-M.}\ \bibnamefont {Wu}}, \bibinfo {author} {\bibfnamefont {Z.-C.}\ \bibnamefont {Chen}}, \ and\ \bibinfo {author} {\bibfnamefont {Q.-G.}\ \bibnamefont {Huang}},\ }\href {\doibase 10.1007/s11433-023-2252-4} {\bibfield  {journal} {\bibinfo  {journal} {Sci. China Phys. Mech. Astron.}\ }\textbf {\bibinfo {volume} {66}},\ \bibinfo {pages} {120402} (\bibinfo {year} {2023})},\ \Eprint {http://arxiv.org/abs/2307.00722} {arXiv:2307.00722 [astro-ph.CO]} \BibitemShut {NoStop}%
\bibitem [{\citenamefont {Ellis}\ \emph {et~al.}(2024{\natexlab{b}})\citenamefont {Ellis}, \citenamefont {Fairbairn}, \citenamefont {H\"utsi}, \citenamefont {Urrutia}, \citenamefont {Vaskonen},\ and\ \citenamefont {Veerm\"ae}}]{Ellis:2024wdh}%
  \BibitemOpen
  \bibfield  {author} {\bibinfo {author} {\bibfnamefont {J.}~\bibnamefont {Ellis}}, \bibinfo {author} {\bibfnamefont {M.}~\bibnamefont {Fairbairn}}, \bibinfo {author} {\bibfnamefont {G.}~\bibnamefont {H\"utsi}}, \bibinfo {author} {\bibfnamefont {J.}~\bibnamefont {Urrutia}}, \bibinfo {author} {\bibfnamefont {V.}~\bibnamefont {Vaskonen}}, \ and\ \bibinfo {author} {\bibfnamefont {H.}~\bibnamefont {Veerm\"ae}},\ }\href@noop {} {\  (\bibinfo {year} {2024}{\natexlab{b}})},\ \Eprint {http://arxiv.org/abs/2403.19650} {arXiv:2403.19650 [astro-ph.CO]} \BibitemShut {NoStop}%
\bibitem [{\citenamefont {Raidal}\ \emph {et~al.}(2024)\citenamefont {Raidal}, \citenamefont {Urrutia}, \citenamefont {Vaskonen},\ and\ \citenamefont {Veerm\"ae}}]{Raidal:2024odr}%
  \BibitemOpen
  \bibfield  {author} {\bibinfo {author} {\bibfnamefont {J.}~\bibnamefont {Raidal}}, \bibinfo {author} {\bibfnamefont {J.}~\bibnamefont {Urrutia}}, \bibinfo {author} {\bibfnamefont {V.}~\bibnamefont {Vaskonen}}, \ and\ \bibinfo {author} {\bibfnamefont {H.}~\bibnamefont {Veerm\"ae}},\ }\href@noop {} {\  (\bibinfo {year} {2024})},\ \Eprint {http://arxiv.org/abs/2406.05125} {arXiv:2406.05125 [astro-ph.CO]} \BibitemShut {NoStop}%
\bibitem [{\citenamefont {Afzal}\ \emph {et~al.}(2023)\citenamefont {Afzal} \emph {et~al.}}]{NANOGrav:2023hvm}%
  \BibitemOpen
  \bibfield  {author} {\bibinfo {author} {\bibfnamefont {A.}~\bibnamefont {Afzal}} \emph {et~al.} (\bibinfo {collaboration} {NANOGrav}),\ }\href {\doibase 10.3847/2041-8213/acdc91} {\bibfield  {journal} {\bibinfo  {journal} {Astrophys. J. Lett.}\ }\textbf {\bibinfo {volume} {951}},\ \bibinfo {pages} {L11} (\bibinfo {year} {2023})},\ \bibinfo {note} {[Erratum: Astrophys.J.Lett. 971, L27 (2024), Erratum: Astrophys.J. 971, L27 (2024)]},\ \Eprint {http://arxiv.org/abs/2306.16219} {arXiv:2306.16219 [astro-ph.HE]} \BibitemShut {NoStop}%
\bibitem [{\citenamefont {Figueroa}\ \emph {et~al.}(2024)\citenamefont {Figueroa}, \citenamefont {Pieroni}, \citenamefont {Ricciardone},\ and\ \citenamefont {Simakachorn}}]{Figueroa:2023zhu}%
  \BibitemOpen
  \bibfield  {author} {\bibinfo {author} {\bibfnamefont {D.~G.}\ \bibnamefont {Figueroa}}, \bibinfo {author} {\bibfnamefont {M.}~\bibnamefont {Pieroni}}, \bibinfo {author} {\bibfnamefont {A.}~\bibnamefont {Ricciardone}}, \ and\ \bibinfo {author} {\bibfnamefont {P.}~\bibnamefont {Simakachorn}},\ }\href {\doibase 10.1103/PhysRevLett.132.171002} {\bibfield  {journal} {\bibinfo  {journal} {Phys. Rev. Lett.}\ }\textbf {\bibinfo {volume} {132}},\ \bibinfo {pages} {171002} (\bibinfo {year} {2024})},\ \Eprint {http://arxiv.org/abs/2307.02399} {arXiv:2307.02399 [astro-ph.CO]} \BibitemShut {NoStop}%
\bibitem [{\citenamefont {Ellis}\ \emph {et~al.}(2024{\natexlab{c}})\citenamefont {Ellis}, \citenamefont {Fairbairn}, \citenamefont {Franciolini}, \citenamefont {H\"utsi}, \citenamefont {Iovino}, \citenamefont {Lewicki}, \citenamefont {Raidal}, \citenamefont {Urrutia}, \citenamefont {Vaskonen},\ and\ \citenamefont {Veerm\"ae}}]{Ellis:2023oxs}%
  \BibitemOpen
  \bibfield  {author} {\bibinfo {author} {\bibfnamefont {J.}~\bibnamefont {Ellis}}, \bibinfo {author} {\bibfnamefont {M.}~\bibnamefont {Fairbairn}}, \bibinfo {author} {\bibfnamefont {G.}~\bibnamefont {Franciolini}}, \bibinfo {author} {\bibfnamefont {G.}~\bibnamefont {H\"utsi}}, \bibinfo {author} {\bibfnamefont {A.}~\bibnamefont {Iovino}}, \bibinfo {author} {\bibfnamefont {M.}~\bibnamefont {Lewicki}}, \bibinfo {author} {\bibfnamefont {M.}~\bibnamefont {Raidal}}, \bibinfo {author} {\bibfnamefont {J.}~\bibnamefont {Urrutia}}, \bibinfo {author} {\bibfnamefont {V.}~\bibnamefont {Vaskonen}}, \ and\ \bibinfo {author} {\bibfnamefont {H.}~\bibnamefont {Veerm\"ae}},\ }\href {\doibase 10.1103/PhysRevD.109.023522} {\bibfield  {journal} {\bibinfo  {journal} {Phys. Rev. D}\ }\textbf {\bibinfo {volume} {109}},\ \bibinfo {pages} {023522} (\bibinfo {year} {2024}{\natexlab{c}})},\ \Eprint {http://arxiv.org/abs/2308.08546} {arXiv:2308.08546 [astro-ph.CO]} \BibitemShut {NoStop}%
\bibitem [{\citenamefont {Mingarelli}\ \emph {et~al.}(2013)\citenamefont {Mingarelli}, \citenamefont {Sidery}, \citenamefont {Mandel},\ and\ \citenamefont {Vecchio}}]{Mingarelli:2013dsa}%
  \BibitemOpen
  \bibfield  {author} {\bibinfo {author} {\bibfnamefont {C.~M.~F.}\ \bibnamefont {Mingarelli}}, \bibinfo {author} {\bibfnamefont {T.}~\bibnamefont {Sidery}}, \bibinfo {author} {\bibfnamefont {I.}~\bibnamefont {Mandel}}, \ and\ \bibinfo {author} {\bibfnamefont {A.}~\bibnamefont {Vecchio}},\ }\href {\doibase 10.1103/PhysRevD.88.062005} {\bibfield  {journal} {\bibinfo  {journal} {Phys. Rev. D}\ }\textbf {\bibinfo {volume} {88}},\ \bibinfo {pages} {062005} (\bibinfo {year} {2013})},\ \Eprint {http://arxiv.org/abs/1306.5394} {arXiv:1306.5394 [astro-ph.HE]} \BibitemShut {NoStop}%
\bibitem [{\citenamefont {Cornish}\ and\ \citenamefont {van Haasteren}(2014)}]{Cornish:2014rva}%
  \BibitemOpen
  \bibfield  {author} {\bibinfo {author} {\bibfnamefont {N.~J.}\ \bibnamefont {Cornish}}\ and\ \bibinfo {author} {\bibfnamefont {R.}~\bibnamefont {van Haasteren}},\ }\href@noop {} {\  (\bibinfo {year} {2014})},\ \Eprint {http://arxiv.org/abs/1406.4511} {arXiv:1406.4511 [gr-qc]} \BibitemShut {NoStop}%
\bibitem [{\citenamefont {Allen}\ \emph {et~al.}(2024)\citenamefont {Allen}, \citenamefont {Agarwal}, \citenamefont {Romano},\ and\ \citenamefont {Valtolina}}]{Allen:2024mtn}%
  \BibitemOpen
  \bibfield  {author} {\bibinfo {author} {\bibfnamefont {B.}~\bibnamefont {Allen}}, \bibinfo {author} {\bibfnamefont {D.}~\bibnamefont {Agarwal}}, \bibinfo {author} {\bibfnamefont {J.~D.}\ \bibnamefont {Romano}}, \ and\ \bibinfo {author} {\bibfnamefont {S.}~\bibnamefont {Valtolina}},\ }\href@noop {} {\  (\bibinfo {year} {2024})},\ \Eprint {http://arxiv.org/abs/2406.16031} {arXiv:2406.16031 [gr-qc]} \BibitemShut {NoStop}%
\bibitem [{\citenamefont {Hotinli}\ \emph {et~al.}(2019)\citenamefont {Hotinli}, \citenamefont {Kamionkowski},\ and\ \citenamefont {Jaffe}}]{Hotinli:2019tpc}%
  \BibitemOpen
  \bibfield  {author} {\bibinfo {author} {\bibfnamefont {S.~C.}\ \bibnamefont {Hotinli}}, \bibinfo {author} {\bibfnamefont {M.}~\bibnamefont {Kamionkowski}}, \ and\ \bibinfo {author} {\bibfnamefont {A.~H.}\ \bibnamefont {Jaffe}},\ }\href {\doibase 10.21105/astro.1904.05348} {\bibfield  {journal} {\bibinfo  {journal} {Open J. Astrophys.}\ }\textbf {\bibinfo {volume} {2}},\ \bibinfo {pages} {8} (\bibinfo {year} {2019})},\ \Eprint {http://arxiv.org/abs/1904.05348} {arXiv:1904.05348 [astro-ph.CO]} \BibitemShut {NoStop}%
\bibitem [{\citenamefont {Taylor}\ and\ \citenamefont {Gair}(2013)}]{Taylor:2013esa}%
  \BibitemOpen
  \bibfield  {author} {\bibinfo {author} {\bibfnamefont {S.~R.}\ \bibnamefont {Taylor}}\ and\ \bibinfo {author} {\bibfnamefont {J.~R.}\ \bibnamefont {Gair}},\ }\href {\doibase 10.1103/PhysRevD.88.084001} {\bibfield  {journal} {\bibinfo  {journal} {Phys. Rev. D}\ }\textbf {\bibinfo {volume} {88}},\ \bibinfo {pages} {084001} (\bibinfo {year} {2013})},\ \Eprint {http://arxiv.org/abs/1306.5395} {arXiv:1306.5395 [gr-qc]} \BibitemShut {NoStop}%
\bibitem [{\citenamefont {Taylor}\ \emph {et~al.}(2020)\citenamefont {Taylor}, \citenamefont {van Haasteren},\ and\ \citenamefont {Sesana}}]{Taylor:2020zpk}%
  \BibitemOpen
  \bibfield  {author} {\bibinfo {author} {\bibfnamefont {S.~R.}\ \bibnamefont {Taylor}}, \bibinfo {author} {\bibfnamefont {R.}~\bibnamefont {van Haasteren}}, \ and\ \bibinfo {author} {\bibfnamefont {A.}~\bibnamefont {Sesana}},\ }\href {\doibase 10.1103/PhysRevD.102.084039} {\bibfield  {journal} {\bibinfo  {journal} {Phys. Rev. D}\ }\textbf {\bibinfo {volume} {102}},\ \bibinfo {pages} {084039} (\bibinfo {year} {2020})},\ \Eprint {http://arxiv.org/abs/2006.04810} {arXiv:2006.04810 [astro-ph.IM]} \BibitemShut {NoStop}%
\bibitem [{\citenamefont {Gardiner}\ \emph {et~al.}(2024)\citenamefont {Gardiner}, \citenamefont {Kelley}, \citenamefont {Lemke},\ and\ \citenamefont {Mitridate}}]{Gardiner:2023zzr}%
  \BibitemOpen
  \bibfield  {author} {\bibinfo {author} {\bibfnamefont {E.~C.}\ \bibnamefont {Gardiner}}, \bibinfo {author} {\bibfnamefont {L.~Z.}\ \bibnamefont {Kelley}}, \bibinfo {author} {\bibfnamefont {A.-M.}\ \bibnamefont {Lemke}}, \ and\ \bibinfo {author} {\bibfnamefont {A.}~\bibnamefont {Mitridate}},\ }\href {\doibase 10.3847/1538-4357/ad2be8} {\bibfield  {journal} {\bibinfo  {journal} {Astrophys. J.}\ }\textbf {\bibinfo {volume} {965}},\ \bibinfo {pages} {164} (\bibinfo {year} {2024})},\ \Eprint {http://arxiv.org/abs/2309.07227} {arXiv:2309.07227 [astro-ph.HE]} \BibitemShut {NoStop}%
\bibitem [{\citenamefont {Sato-Polito}\ and\ \citenamefont {Kamionkowski}(2024)}]{Sato-Polito:2023spo}%
  \BibitemOpen
  \bibfield  {author} {\bibinfo {author} {\bibfnamefont {G.}~\bibnamefont {Sato-Polito}}\ and\ \bibinfo {author} {\bibfnamefont {M.}~\bibnamefont {Kamionkowski}},\ }\href {\doibase 10.1103/PhysRevD.109.123544} {\bibfield  {journal} {\bibinfo  {journal} {Phys. Rev. D}\ }\textbf {\bibinfo {volume} {109}},\ \bibinfo {pages} {123544} (\bibinfo {year} {2024})},\ \Eprint {http://arxiv.org/abs/2305.05690} {arXiv:2305.05690 [astro-ph.CO]} \BibitemShut {NoStop}%
\bibitem [{\citenamefont {Sah}\ \emph {et~al.}(2024)\citenamefont {Sah}, \citenamefont {Mukherjee}, \citenamefont {Saeedzadeh}, \citenamefont {Babul}, \citenamefont {Tremmel},\ and\ \citenamefont {Quinn}}]{Sah:2024oyg}%
  \BibitemOpen
  \bibfield  {author} {\bibinfo {author} {\bibfnamefont {M.~R.}\ \bibnamefont {Sah}}, \bibinfo {author} {\bibfnamefont {S.}~\bibnamefont {Mukherjee}}, \bibinfo {author} {\bibfnamefont {V.}~\bibnamefont {Saeedzadeh}}, \bibinfo {author} {\bibfnamefont {A.}~\bibnamefont {Babul}}, \bibinfo {author} {\bibfnamefont {M.}~\bibnamefont {Tremmel}}, \ and\ \bibinfo {author} {\bibfnamefont {T.~R.}\ \bibnamefont {Quinn}},\ }\href {\doibase 10.1093/mnras/stae1930} {\bibfield  {journal} {\bibinfo  {journal} {Mon. Not. Roy. Astron. Soc.}\ }\textbf {\bibinfo {volume} {533}},\ \bibinfo {pages} {1568} (\bibinfo {year} {2024})},\ \Eprint {http://arxiv.org/abs/2404.14508} {arXiv:2404.14508 [astro-ph.CO]} \BibitemShut {NoStop}%
\bibitem [{\citenamefont {Agazie}\ \emph {et~al.}(2023{\natexlab{c}})\citenamefont {Agazie} \emph {et~al.}}]{NANOGrav:2023tcn}%
  \BibitemOpen
  \bibfield  {author} {\bibinfo {author} {\bibfnamefont {G.}~\bibnamefont {Agazie}} \emph {et~al.} (\bibinfo {collaboration} {NANOGrav}),\ }\href {\doibase 10.3847/2041-8213/acf4fd} {\bibfield  {journal} {\bibinfo  {journal} {Astrophys. J. Lett.}\ }\textbf {\bibinfo {volume} {956}},\ \bibinfo {pages} {L3} (\bibinfo {year} {2023}{\natexlab{c}})},\ \Eprint {http://arxiv.org/abs/2306.16221} {arXiv:2306.16221 [astro-ph.HE]} \BibitemShut {NoStop}%
\bibitem [{\citenamefont {Pol}\ \emph {et~al.}(2022)\citenamefont {Pol}, \citenamefont {Taylor},\ and\ \citenamefont {Romano}}]{Pol:2022sjn}%
  \BibitemOpen
  \bibfield  {author} {\bibinfo {author} {\bibfnamefont {N.}~\bibnamefont {Pol}}, \bibinfo {author} {\bibfnamefont {S.~R.}\ \bibnamefont {Taylor}}, \ and\ \bibinfo {author} {\bibfnamefont {J.~D.}\ \bibnamefont {Romano}},\ }\href {\doibase 10.3847/1538-4357/ac9836} {\bibfield  {journal} {\bibinfo  {journal} {Astrophys. J.}\ }\textbf {\bibinfo {volume} {940}},\ \bibinfo {pages} {173} (\bibinfo {year} {2022})},\ \Eprint {http://arxiv.org/abs/2206.09936} {arXiv:2206.09936 [astro-ph.HE]} \BibitemShut {NoStop}%
\bibitem [{\citenamefont {Lemke}\ \emph {et~al.}(2024)\citenamefont {Lemke}, \citenamefont {Mitridate},\ and\ \citenamefont {Gersbach}}]{Lemke:2024cdu}%
  \BibitemOpen
  \bibfield  {author} {\bibinfo {author} {\bibfnamefont {A.-M.}\ \bibnamefont {Lemke}}, \bibinfo {author} {\bibfnamefont {A.}~\bibnamefont {Mitridate}}, \ and\ \bibinfo {author} {\bibfnamefont {K.~A.}\ \bibnamefont {Gersbach}},\ }\href@noop {} {\  (\bibinfo {year} {2024})},\ \Eprint {http://arxiv.org/abs/2407.08705} {arXiv:2407.08705 [astro-ph.HE]} \BibitemShut {NoStop}%
\bibitem [{\citenamefont {Depta}\ \emph {et~al.}(2024)\citenamefont {Depta}, \citenamefont {Domcke}, \citenamefont {Franciolini},\ and\ \citenamefont {Pieroni}}]{Depta:2024ykq}%
  \BibitemOpen
  \bibfield  {author} {\bibinfo {author} {\bibfnamefont {P.~F.}\ \bibnamefont {Depta}}, \bibinfo {author} {\bibfnamefont {V.}~\bibnamefont {Domcke}}, \bibinfo {author} {\bibfnamefont {G.}~\bibnamefont {Franciolini}}, \ and\ \bibinfo {author} {\bibfnamefont {M.}~\bibnamefont {Pieroni}},\ }\href@noop {} {\  (\bibinfo {year} {2024})},\ \Eprint {http://arxiv.org/abs/2407.14460} {arXiv:2407.14460 [astro-ph.CO]} \BibitemShut {NoStop}%
\bibitem [{\citenamefont {Ellis}\ \emph {et~al.}(2023)\citenamefont {Ellis}, \citenamefont {Fairbairn}, \citenamefont {H\"utsi}, \citenamefont {Raidal}, \citenamefont {Urrutia}, \citenamefont {Vaskonen},\ and\ \citenamefont {Veerm\"ae}}]{Ellis:2023owy}%
  \BibitemOpen
  \bibfield  {author} {\bibinfo {author} {\bibfnamefont {J.}~\bibnamefont {Ellis}}, \bibinfo {author} {\bibfnamefont {M.}~\bibnamefont {Fairbairn}}, \bibinfo {author} {\bibfnamefont {G.}~\bibnamefont {H\"utsi}}, \bibinfo {author} {\bibfnamefont {M.}~\bibnamefont {Raidal}}, \bibinfo {author} {\bibfnamefont {J.}~\bibnamefont {Urrutia}}, \bibinfo {author} {\bibfnamefont {V.}~\bibnamefont {Vaskonen}}, \ and\ \bibinfo {author} {\bibfnamefont {H.}~\bibnamefont {Veerm\"ae}},\ }\href {\doibase 10.1051/0004-6361/202346268} {\bibfield  {journal} {\bibinfo  {journal} {Astron. Astrophys.}\ }\textbf {\bibinfo {volume} {676}},\ \bibinfo {pages} {A38} (\bibinfo {year} {2023})},\ \Eprint {http://arxiv.org/abs/2301.13854} {arXiv:2301.13854 [astro-ph.CO]} \BibitemShut {NoStop}%
\bibitem [{\citenamefont {Sato-Polito}\ and\ \citenamefont {Zaldarriaga}(2024)}]{Sato-Polito:2024lew}%
  \BibitemOpen
  \bibfield  {author} {\bibinfo {author} {\bibfnamefont {G.}~\bibnamefont {Sato-Polito}}\ and\ \bibinfo {author} {\bibfnamefont {M.}~\bibnamefont {Zaldarriaga}},\ }\href@noop {} {\  (\bibinfo {year} {2024})},\ \Eprint {http://arxiv.org/abs/2406.17010} {arXiv:2406.17010 [astro-ph.CO]} \BibitemShut {NoStop}%
\bibitem [{\citenamefont {Xue}\ \emph {et~al.}(2024)\citenamefont {Xue}, \citenamefont {Pan},\ and\ \citenamefont {Dai}}]{Xue:2024qtx}%
  \BibitemOpen
  \bibfield  {author} {\bibinfo {author} {\bibfnamefont {X.}~\bibnamefont {Xue}}, \bibinfo {author} {\bibfnamefont {Z.}~\bibnamefont {Pan}}, \ and\ \bibinfo {author} {\bibfnamefont {L.}~\bibnamefont {Dai}},\ }\href@noop {} {\  (\bibinfo {year} {2024})},\ \Eprint {http://arxiv.org/abs/2409.19516} {arXiv:2409.19516 [astro-ph.CO]} \BibitemShut {NoStop}%
\bibitem [{\citenamefont {Enoki}\ and\ \citenamefont {Nagashima}(2007)}]{Enoki:2006kj}%
  \BibitemOpen
  \bibfield  {author} {\bibinfo {author} {\bibfnamefont {M.}~\bibnamefont {Enoki}}\ and\ \bibinfo {author} {\bibfnamefont {M.}~\bibnamefont {Nagashima}},\ }\href {\doibase 10.1143/PTP.117.241} {\bibfield  {journal} {\bibinfo  {journal} {Prog. Theor. Phys.}\ }\textbf {\bibinfo {volume} {117}},\ \bibinfo {pages} {241} (\bibinfo {year} {2007})},\ \Eprint {http://arxiv.org/abs/astro-ph/0609377} {arXiv:astro-ph/0609377} \BibitemShut {NoStop}%
\bibitem [{\citenamefont {Sesana}(2013)}]{Sesana:2013wja}%
  \BibitemOpen
  \bibfield  {author} {\bibinfo {author} {\bibfnamefont {A.}~\bibnamefont {Sesana}},\ }\href {\doibase 10.1088/0264-9381/30/22/224014} {\bibfield  {journal} {\bibinfo  {journal} {Class. Quant. Grav.}\ }\textbf {\bibinfo {volume} {30}},\ \bibinfo {pages} {224014} (\bibinfo {year} {2013})},\ \Eprint {http://arxiv.org/abs/1307.2600} {arXiv:1307.2600 [astro-ph.CO]} \BibitemShut {NoStop}%
\bibitem [{\citenamefont {Press}\ and\ \citenamefont {Schechter}(1974)}]{Press:1973iz}%
  \BibitemOpen
  \bibfield  {author} {\bibinfo {author} {\bibfnamefont {W.~H.}\ \bibnamefont {Press}}\ and\ \bibinfo {author} {\bibfnamefont {P.}~\bibnamefont {Schechter}},\ }\href {\doibase 10.1086/152650} {\bibfield  {journal} {\bibinfo  {journal} {Astrophys. J.}\ }\textbf {\bibinfo {volume} {187}},\ \bibinfo {pages} {425} (\bibinfo {year} {1974})}\BibitemShut {NoStop}%
\bibitem [{\citenamefont {Bond}\ \emph {et~al.}(1991)\citenamefont {Bond}, \citenamefont {Cole}, \citenamefont {Efstathiou},\ and\ \citenamefont {Kaiser}}]{Bond:1990iw}%
  \BibitemOpen
  \bibfield  {author} {\bibinfo {author} {\bibfnamefont {J.~R.}\ \bibnamefont {Bond}}, \bibinfo {author} {\bibfnamefont {S.}~\bibnamefont {Cole}}, \bibinfo {author} {\bibfnamefont {G.}~\bibnamefont {Efstathiou}}, \ and\ \bibinfo {author} {\bibfnamefont {N.}~\bibnamefont {Kaiser}},\ }\href {\doibase 10.1086/170520} {\bibfield  {journal} {\bibinfo  {journal} {Astrophys. J.}\ }\textbf {\bibinfo {volume} {379}},\ \bibinfo {pages} {440} (\bibinfo {year} {1991})}\BibitemShut {NoStop}%
\bibitem [{\citenamefont {{Lacey}}\ and\ \citenamefont {{Cole}}(1993)}]{1993MNRAS.262..627L}%
  \BibitemOpen
  \bibfield  {author} {\bibinfo {author} {\bibfnamefont {C.}~\bibnamefont {{Lacey}}}\ and\ \bibinfo {author} {\bibfnamefont {S.}~\bibnamefont {{Cole}}},\ }\href {\doibase 10.1093/mnras/262.3.627} {\bibfield  {journal} {\bibinfo  {journal} {\mnras}\ }\textbf {\bibinfo {volume} {262}},\ \bibinfo {pages} {627} (\bibinfo {year} {1993})}\BibitemShut {NoStop}%
\bibitem [{\citenamefont {Girelli}\ \emph {et~al.}(2020)\citenamefont {Girelli}, \citenamefont {Pozzetti}, \citenamefont {Bolzonella}, \citenamefont {Giocoli}, \citenamefont {Marulli},\ and\ \citenamefont {Baldi}}]{Girelli:2020goz}%
  \BibitemOpen
  \bibfield  {author} {\bibinfo {author} {\bibfnamefont {G.}~\bibnamefont {Girelli}}, \bibinfo {author} {\bibfnamefont {L.}~\bibnamefont {Pozzetti}}, \bibinfo {author} {\bibfnamefont {M.}~\bibnamefont {Bolzonella}}, \bibinfo {author} {\bibfnamefont {C.}~\bibnamefont {Giocoli}}, \bibinfo {author} {\bibfnamefont {F.}~\bibnamefont {Marulli}}, \ and\ \bibinfo {author} {\bibfnamefont {M.}~\bibnamefont {Baldi}},\ }\href {\doibase 10.1051/0004-6361/201936329} {\bibfield  {journal} {\bibinfo  {journal} {Astron. Astrophys.}\ }\textbf {\bibinfo {volume} {634}},\ \bibinfo {pages} {A135} (\bibinfo {year} {2020})},\ \Eprint {http://arxiv.org/abs/2001.02230} {arXiv:2001.02230 [astro-ph.CO]} \BibitemShut {NoStop}%
\bibitem [{\citenamefont {Peters}\ and\ \citenamefont {Mathews}(1963)}]{Peters:1963ux}%
  \BibitemOpen
  \bibfield  {author} {\bibinfo {author} {\bibfnamefont {P.~C.}\ \bibnamefont {Peters}}\ and\ \bibinfo {author} {\bibfnamefont {J.}~\bibnamefont {Mathews}},\ }\href {\doibase 10.1103/PhysRev.131.435} {\bibfield  {journal} {\bibinfo  {journal} {Phys. Rev.}\ }\textbf {\bibinfo {volume} {131}},\ \bibinfo {pages} {435} (\bibinfo {year} {1963})}\BibitemShut {NoStop}%
\bibitem [{\citenamefont {Semenzato}\ \emph {et~al.}(2024)\citenamefont {Semenzato}, \citenamefont {Casey-Clyde}, \citenamefont {Mingarelli}, \citenamefont {Raccanelli}, \citenamefont {Bellomo}, \citenamefont {Bartolo},\ and\ \citenamefont {Bertacca}}]{Semenzato:2024mtn}%
  \BibitemOpen
  \bibfield  {author} {\bibinfo {author} {\bibfnamefont {F.}~\bibnamefont {Semenzato}}, \bibinfo {author} {\bibfnamefont {J.~A.}\ \bibnamefont {Casey-Clyde}}, \bibinfo {author} {\bibfnamefont {C.~M.~F.}\ \bibnamefont {Mingarelli}}, \bibinfo {author} {\bibfnamefont {A.}~\bibnamefont {Raccanelli}}, \bibinfo {author} {\bibfnamefont {N.}~\bibnamefont {Bellomo}}, \bibinfo {author} {\bibfnamefont {N.}~\bibnamefont {Bartolo}}, \ and\ \bibinfo {author} {\bibfnamefont {D.}~\bibnamefont {Bertacca}},\ }\href@noop {} {\  (\bibinfo {year} {2024})},\ \Eprint {http://arxiv.org/abs/2411.00532} {arXiv:2411.00532 [astro-ph.CO]} \BibitemShut {NoStop}%
\bibitem [{\citenamefont {Agazie}\ \emph {et~al.}(2023{\natexlab{d}})\citenamefont {Agazie} \emph {et~al.}}]{NANOGrav:2023pdq}%
  \BibitemOpen
  \bibfield  {author} {\bibinfo {author} {\bibfnamefont {G.}~\bibnamefont {Agazie}} \emph {et~al.} (\bibinfo {collaboration} {NANOGrav}),\ }\href {\doibase 10.3847/2041-8213/ace18a} {\bibfield  {journal} {\bibinfo  {journal} {Astrophys. J. Lett.}\ }\textbf {\bibinfo {volume} {951}},\ \bibinfo {pages} {L50} (\bibinfo {year} {2023}{\natexlab{d}})},\ \Eprint {http://arxiv.org/abs/2306.16222} {arXiv:2306.16222 [astro-ph.HE]} \BibitemShut {NoStop}%
\end{thebibliography}%

\end{document}